\documentclass[useAMS,usenatbib]{mnras}
\pdfoutput=1
\usepackage{times}
\usepackage{graphicx}
\usepackage{epsfig}
\usepackage[T1]{fontenc}
\usepackage{ifpdf}
\usepackage[varg]{txfonts}
\usepackage{hyperref}

\newcommand{\Msun}      {\mbox{\,$M_{\mathord\odot}$}}

\title[Classifying IGR~J15038--6021 as a magnetic CV with a massive white dwarf]{Classifying IGR~J15038--6021 as a magnetic CV with a massive white dwarf}

\author[Tomsick et al.]{
John A. Tomsick$^{1}$\thanks{E-mail: jtomsick@berkeley.edu (JAT)}, Snehaa Ganesh Kumar$^{2}$, Benjamin M. Coughenour$^{1}$,
\newauthor
Aarran W. Shaw$^{3}$, Koji Mukai$^{4,5}$, Jeremy Hare$^{6,7,8}$, Ma\"{i}ca Clavel$^{9}$, Roman Krivonos$^{10}$, 
\newauthor
Francesca M. Fornasini$^{11}$, Julian Gerber$^{12}$, and Alyson Joens$^{1}$\\
$^{1}$Space Sciences Laboratory, 7 Gauss Way, University of California, Berkeley, CA 94720-7450, USA\\
$^{2}$Department of Astronomy, University of California, Berkeley, CA 94720, USA\\
$^{3}$Department of Physics, University of Nevada, Reno, NV 89557, USA\\
$^{4}$CRESST and X-ray Astrophysics Laboratory, NASA Goddard Space Flight Center, Greenbelt, MD 20771, USA\\
$^{5}$Department of Physics, University of Maryland, Baltimore County, 1000 Hilltop Circle, Baltimore, MD 21250, USA\\
$^{6}$NASA Goddard Space Flight Center, Greenbelt, MD 20771, USA\\
$^{7}$Center for Research and Exploration in Space Science and Technology, NASA/GSFC, Greenbelt, Maryland 20771, USA\\
$^{8}$The Catholic University of America, 620 Michigan Ave., N.E. Washington, DC 20064, USA\\
$^{9}$Universit\'{e} Grenoble Alpes, CNRS, IPAG, F38000 Grenoble, France \\
$^{10}$Space Research Institute, Russian Academy of Sciences, Profsoyuznaya 84/32, 117997 Moscow, Russia\\
$^{11}$Stonehill College, 320 Washington Street, Easton, MA 02357, USA\\
$^{12}$Columbia Astrophysics Laboratory, Columbia University, New York, NY 10027, USA\\
}

\begin{document}


\def\lsim{\mathrel{\lower .85ex\hbox{\rlap{$\sim$}\raise
.95ex\hbox{$<$} }}}
\def\gsim{\mathrel{\lower .80ex\hbox{\rlap{$\sim$}\raise
.90ex\hbox{$>$} }}}

\pagerange{\pageref{firstpage}--\pageref{lastpage}} \pubyear{2014}

\maketitle

\label{firstpage}

\begin{abstract}

\noindent
Cataclysmic variables (CVs) are binary systems consisting of a white dwarf (WD) accreting matter from a companion star.  Observations of CVs provide an opportunity to learn about accretion disks, the physics of compact objects, classical novae, and the evolution of the binary and the WD that may ultimately end in a type Ia supernova (SN).  As type Ia SNe involve a WD reaching the Chandrasekhar limit or merging WDs, WD mass measurements are particularly important for elucidating the path from CV to type Ia SN.  For intermediate polar (IP) type CVs, the WD mass is related to the bremsstrahlung temperature of material in the accretion column, which typically peaks at X-ray energies.  Thus, the IPs with the strongest hard X-ray emission, such as those discovered by the {\em INTEGRAL} satellite, are expected to have the highest masses. Here, we report on {\em XMM-Newton}, {\em NuSTAR}, and optical observations of IGR~J15038--6021.  We find an X-ray periodicity of $1678\pm 2$\,s, which we interpret as the WD spin period.  From fitting the 0.3-79\,keV spectrum with a model that uses the relationship between the WD mass and the post-shock temperature, we measure a WD mass of $1.36^{+0.04}_{-0.11}$\Msun.  This follows an earlier study of IGR~J14091--6108, which also has a WD with a mass approaching the Chandrasekhar limit.  We demonstrate that these are both outliers among IPs in having massive WDs and discuss the results in the context of WD mass studies as well as the implications for WD mass evolution.

\end{abstract}


\section{Introduction}

The {\em International Gamma-ray Astrophysics Laboratory (INTEGRAL)} satellite has been used to study the $>$20\,keV sky since it began operations in 2002.  Its survey of the sky has led to the detection of over a thousand hard X-ray sources \citep{bird16,krivonos17,krivonos22}.  The sources include active galactic nuclei (AGN), X-ray binaries, pulsar wind nebulae, cataclysmic variables (CVs), and other source types.  While many of these sources were previously known, a large fraction of them were unknown or poorly studied prior to {\em INTEGRAL}, and these are the {\em INTEGRAL} gamma-ray (IGR) sources.  For most of the IGR sources, dedicated follow-up observations (e.g., X-ray, optical, near-IR) are necessary to classify them.  \cite{krivonos21} provides a recent review of the {\em INTEGRAL} survey and describes follow-up efforts, including X-ray observations to localize the sources and identify optical or near-IR counterparts.  Most of the CVs that have been detected by {\em INTEGRAL} are magnetic CVs of intermediate polar (IP) type.  \cite{lutovinov20} lists 78 CVs (IGR and non-IGR) detected by {\em INTEGRAL}, and 51 are IPs or IP candidates.  Also, as of 2021 December, the Intermediate Polars website\footnote{See \href{https://asd.gsfc.nasa.gov/Koji.Mukai/iphome/iphome.html}{https://asd.gsfc.nasa.gov/Koji.Mukai/iphome/iphome.html}} has listed 150 known IPs or IP candidates, of which 33 are IGR sources.  Thus, {\em INTEGRAL} has significantly increased the population of known IPs.

This work focuses on follow-up observations of one particular source, IGR~J15038--6021 (J15038), which was first reported by \cite{bird16}, and initial follow-up X-ray measurements with the {\em Neil Gehrels Swift} satellite were presented in \cite{landi17}.  Prior to this work, we carried out observations of IGR sources with the {\em Chandra X-ray Observatory} to localize the sources, allowing us to identify counterparts at other wavelengths.  {\em Chandra} also provides information about the soft X-ray spectrum, and we studied J15038 as part of a {\em Chandra} study of 15 IGR sources \citep{tomsick20}.  J15038 and IGR~J18007--4146 (J18007) were the two sources with {\em Gaia} optical counterparts and parallax distance measurements, and we classified both of these as likely IPs based on their X-ray luminosities, X-ray spectra, and near-IR magnitudes.  Follow-up X-ray observations of J18007 led to the discovery of an X-ray periodicity at $422.8\pm 0.7$\,s and strong Fe emission lines, confirming its IP nature \citep{coughenour22}. 

J15038 was identified with the {\em Chandra} source CXOU~J150415.7--602123 and the VISTA source VVV~J150415.72--602122.87 \citep{tomsick20}.  Based on the {\em Gaia} DR2\footnote{See Section 3.4 for an analysis using the updated EDR3/DR3 distance.} distance of $1.1^{+1.5}_{-0.4}$\,kpc and the {\em Chandra}+{\em INTEGRAL} spectrum, we derived a 0.3-100\,keV luminosity of $(1.6^{+4.4}_{-1.2})\times 10^{33}$\,erg\,s$^{-1}$.  A fit to the {\em Chandra}+{\em INTEGRAL} spectrum with a cutoff power-law model yielded a very hard photon index of $\Gamma$ = --$0.3^{+0.6}_{-0.3}$.  This information, combined with the near-IR magnitudes favoring a stellar type in the K5V-F5V range, motivated the likely IP classification \citep{tomsick20}.

In IPs, the accretion disk is disrupted by the white dwarf's (WD's) magnetic field, and the accreting material is funneled onto the magnetic poles of the WD.  The hard X-rays are produced by shock-heating in the accretion column, and the measured temperature depends on the mass of the WD \citep{srr05,suleimanov16}.  By obtaining high-quality hard X-ray spectra, it is possible to measure the maximum temperature of the material in the accretion column and to constrain the WD mass.  The best place to look for high-mass WDs is among CVs with the highest temperatures and the hardest spectra, making CVs found in the 17-100\,keV {\em INTEGRAL} Galactic Plane survey \citep{bird16,krivonos17,krivonos22} excellent candidates for hosting more massive WDs.  Much of our effort has focused on IGR sources, but the {Neil Gehrels Swift} Burst Alert Transient observations have also been used for studies of hard X-rays from CVs \citep[e.g.,][]{suleimanov22}.

One reason for interest in massive WDs is the question of whether the progenitors of type Ia SNe are merging WDs (the double degenerate, DD, channel) or accreting WDs that detonate when they reach the Chandrasekhar limit (the single degenerate, SD, channel). Based on the relatively low soft X-ray luminosities from nearby elliptical galaxies and galaxy bulges, \cite{gb10} and \cite{distefano10a} argue that $<$5\% of type Ia SNe are from accreting WDs.  Also, pre-explosion {\em Chandra} and {\em Hubble Space Telescope} imaging of individual nearby type Ia SNe place constraints on accreting and nuclear-burning WDs as the SN progenitors \citep{liu12,nielsen12,nielsen13b,nielsen14,graur14}.  The observations provide limits on the luminosity and temperature of the pre-SN WD.  However, the fact that the X-rays can be attenuated by stellar winds, accretion winds, or WD atmospheres \citep{distefano10b,nielsen13a} means that the accreting WD scenario remains a possibility, and the question of the dominance of the SD or DD channels and the detailed physics of type Ia SN detonations is an active area of observational and theoretical work \citep[e.g.,][]{caiazzo21,kilic21,bravo22,neopane22,leising22}.

Even beyond the question of whether CVs are the progenitors of type Ia SNe, there are other open questions for which the WD mass distribution is relevant.  While WDs gain mass during accretion phases, they also have significant mass loss during thermonuclear runaways that result in classical novae.  Determining the mass loss during nova events is related to understanding element formation.  Recent simulation results suggest that while novae make a very significant contribution to the formation of lithium in the Galaxy, WD masses should increase during the accretion-outburst-accretion cycle \citep{starrfield20}.  The WD mass measurements made to date provide evidence that the WDs in CVs are, on average, more massive than WDs that have not undergone mass accretion from a companion \citep{zsg11,shaw20}, which is consistent with the simulation results.

The {\em Nuclear Spectroscopic Telescope Array} \citep[{\em NuSTAR},][]{harrison13} has already provided some interesting hints that average WD masses depend on the age of the population, which would imply a longer accretion duration.  Before {\em NuSTAR}, there were studies of unresolved hard X-ray sources in the Galactic ridge to determine an overall Galactic average WD mass between 0.50\Msun~and 0.66\Msun~\citep{srr05,krivonos07_cv,yuasa12}.  {\em NuSTAR} measurements of individual CVs in the relatively young population in the Norma spiral arm suggest that the temperatures (and therefore masses) may be lower than the masses in the Galactic center region \citep{hong16,fornasini17}.  Also, there is unresolved hard X-ray emission very close to Sgr~A* that may be from IPs with even more massive WDs \citep{perez15,hailey16}.  Studies of individual IPs have also been carried out using the less sensitive Swift/BAT data \citep{suleimanov19}.

A sample of IPs has been observed by {\em NuSTAR}, and the resulting WD mass distribution is shown in \cite{shaw20}.  While \cite{shaw20} did not obtain measurements of WD masses greater than 1.1\Msun, there are other cases where higher masses have been reported.  One example is IGR~J14091--6108 (J14091), which was selected based on its hard X-ray spectrum for a detailed study with {\em NuSTAR} and the {\em X-ray Multi-Mirror Mission} ({\em XMM-Newton} or {\em XMM}), and a constraint of $>$1.3\Msun~was obtained \citep{tomsick16b}.  In addition, a review of hard X-ray IPs includes five more IPs with best estimates of WD mass in excess of 1.1\Msun~\citep{demartino20}.

For this work, we carried out a joint {\em XMM} and {\em NuSTAR} observation of J15038 as well as optical photometry taken at Las Cumbres Observatory (LCO), and the results are reported below.  Section 2 provides information about the observations that were obtained and the data analysis methods.  The results are presented in Section 3, including timing and spectral analysis with {\em XMM}'s large effective area, sensitivity, and soft X-ray (0.3-12\,keV) coverage and {\em NuSTAR}'s sensitive hard X-ray (3-79\,keV) coverage, which is critical for the WD mass measurements.  In addition, we use the LCO observations to search for potential periodic optical signals. A discussion of the results is presented in Section 4.

\section{Observations and Data Analysis}

We analyzed X-ray observations of J15038 made with {\em XMM} on 2020 July 30 and with {\em NuSTAR} on 2020 July 30-31 (Table~\ref{tab:obs}).  The details of the X-ray data reduction are described in Sections 2.1 and 2.2.  The LCO observations occurred on 2022 May 2 and are described in Section 2.3.

\subsection{{\em XMM}}

For the EPIC/pn \citep{struder01} and EPIC/MOS \citep{turner01} instruments, we reduced the data using the {\em XMM} Science Analysis Software v18.0.0 to make images, light curves, and spectra.  All three instruments (pn, MOS1, and MOS2) were in Full Frame mode with the medium thickness optical blocking filter.  We used the standard analysis procedures provided on-line\footnote{See https://www.cosmos.esa.int/web/xmm-newton/sas-threads}.  To check the pn for high levels of particle background, we made a 10-12\,keV light curve with 100\,s time bins using photons from the entire field of view.  At no point does the light curve exceed the high-background threshold of 0.4\,c/s.  However, there is a 600\,s time period near the middle of the observation where the full-field pn count rate drops to zero.  The data report indicates that telemetry glitches occurred for this observation (ObsID 0870790101), which is the cause of the gap. We applied a filter to the data to remove this time segment, and the only impact is a reduction of the exposure time.

\begin{figure*}
\begin{center}
    \includegraphics[width=15cm]{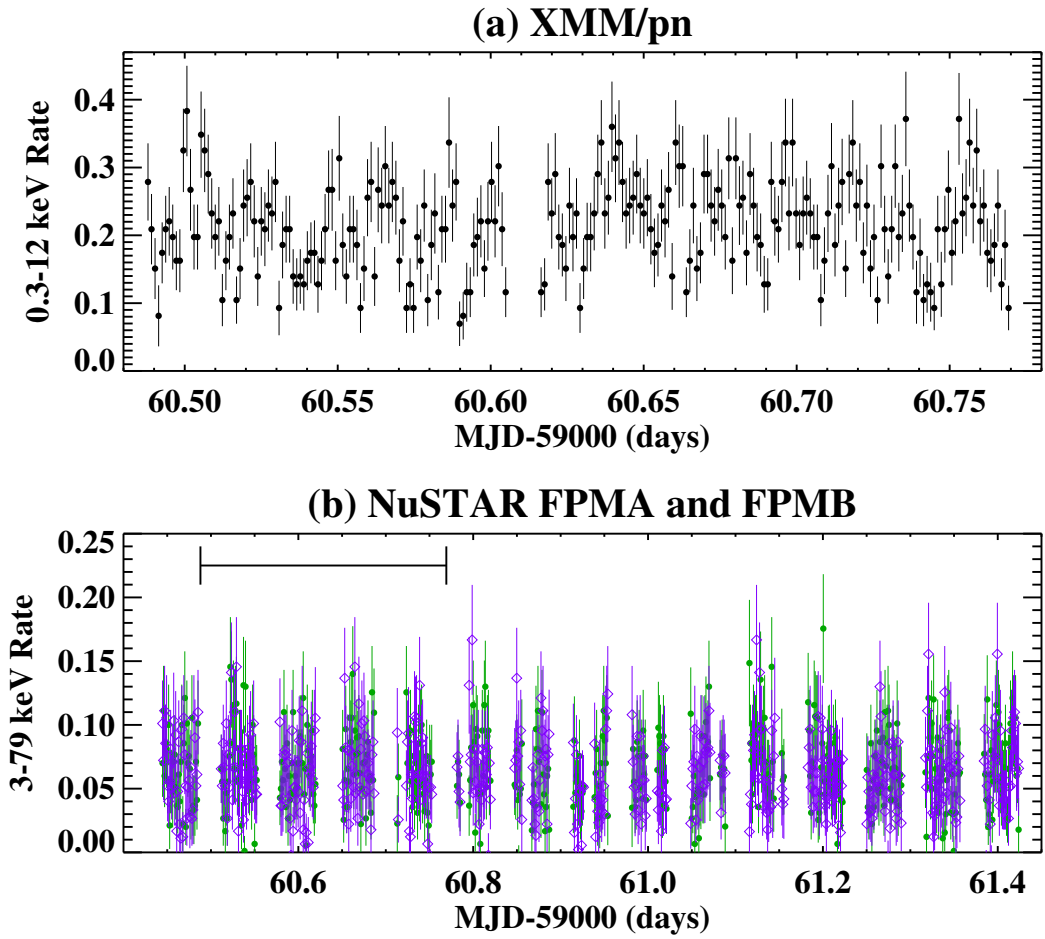}
    \caption{{\em (a)} {\em XMM}/pn 0.3-12\,keV light curve with 100\,s time bins. {\em (b)} {\em NuSTAR} 3-79\,keV light curve for FPMA with green filled circles and for FPMB with purple diamonds.  The time period of the {\em XMM} coverage is marked.
    \label{fig:lc}}
\end{center}
\end{figure*}

The pn and MOS images show a point-like source at the location of J15038 known from the previous {\em Chandra} and {\em Swift} observations.  We extracted source counts from a circular region with a radius of $25^{\prime\prime}$ centered on J15038 and estimated the background using a larger ($50^{\prime\prime}$ radius for pn and $100^{\prime\prime}$ for MOS) circular region on parts of the detectors with no sources\footnote{For pn, the background region was centered at R.A. = 15h 04m 19.4s, Decl. = --60$^{\circ}$ 22$^{\prime}$ 48$^{\prime\prime}$.  For the MOS units, the background region was cetered at R.A. = 15h 04m 41.0s, Decl. = --60$^{\circ}$ 18$^{\prime}$ 37$^{\prime\prime}$.}.  The background subtracted 0.3-12\,keV {\em XMM}/pn count rate is $0.210\pm 0.003$\,c/s, and the light curve is shown in Figure~\ref{fig:lc}a.

A second moderately bright source is detected.  It is relatively close to J15038, with an angular separation between the two sources of $1.45^{\prime}$.  It also appears to be point-like, and there is no reason to suspect that it is related to J15038.  Using the MOS1 data, we determined the source position to be R.A. = 15h 04m 07.86s, Decl. = --60$^{\circ}$ 22$^{\prime}$ 27.2$^{\prime\prime}$ (J2000), and here, we report the discovery of XMMU~J150407.8--602227 (see the Appendix for additional information about this nearby source). 

\subsection{{\em NuSTAR}}

For {\em NuSTAR} \citep{harrison13}, we reduced the data using HEASOFT v6.29b, which includes NUSTARDAS v2.1.1.  The calibration files are from the 2021 October 20 version of the calibration database (CALDB). We ran {\ttfamily nupipeline} to produce event lists for the two {\em NuSTAR} instruments, focal plane modules A and B (FPMA and FPMB), and extracted light curves and spectra with {\ttfamily nuproducts} using a circular source region with a radius of $50''$ and a circular background region with a radius of $75''$. Figure~\ref{fig:image} shows the regions used on the FPMA image.  {\em NuSTAR}'s angular resolution is sufficient to cleanly separate J15038 from XMMU~J150407.8--602227. For J15038, the average 3--79\,keV source count rates are $0.0658\pm 0.0013$ and $0.0617\pm 0.0013$\,c/s for FPMA and FPMB, respectively, and the light curves are shown in Figure~\ref{fig:lc}b.

\subsection{LCO}

The J15038 field was observed with one of the three LCO 1m telescopes located at the South African Astronomical Observatory in Sutherland, South Africa on 2022 May 2 under proposal ID NSF2022A-013. We obtained a total of 228 $r'$-band images with the Sinistro camera operating in the central 2k x 2k window readout mode, enabling readout times of $\approx10$s. Each image had an exposure time of 51s. Bias, dark and sky flat calibration frames were obtained by LCO between 2022 April 30 and 2022 May 1 as part of the standard calibration plan and data reduction was performed in (near) real time by the LCO Beautiful Algorithms to Normalize Zillions of Astronomical Images (BANZAI) pipeline \citep{mccully18}. In addition to standard processing, BANZAI also astrometrically solved individual images using Astrometry.net \citep{lang10}.

\begin{figure*}
\begin{center}
    \includegraphics[width=15cm]{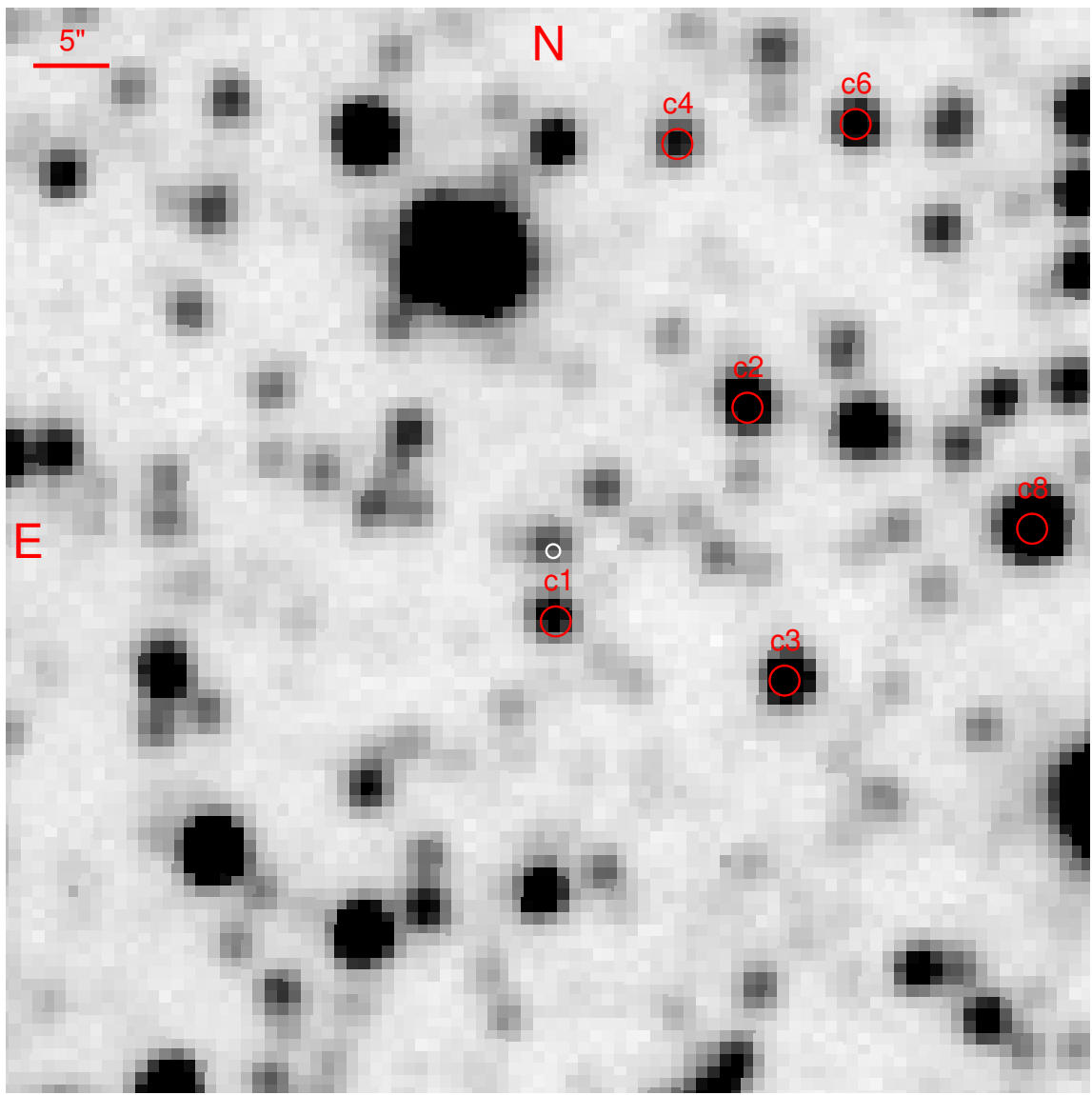}
    \caption{LCO Sinistro $r'$-band average combined image of the J15038 field. The {\em Chandra} position of J15038 \citep[with $0.45''$ uncertainty,][]{tomsick20} is shown as a white circle. The positions of some of the comparison stars we used for photometric calibration are shown as red circles.\label{fig:LCO_image}}
\end{center}
\end{figure*}

The average combined LCO image is presented in Figure~\ref{fig:LCO_image} and shows the {\em Chandra} position and associated uncertainty from \citet{tomsick20}. The {\em Chandra} error circle lies on top of a single source consistent with the optical counterpart identified in {\em Gaia} DR2 by \citet{tomsick20}.\footnote{{\em Gaia} source ID 5876459780108921216} The field surrounding the optical counterpart to J15038 is moderately crowded, such that it was necessary to perform PSF photometry to extract the $r'$-band light curve of the source. We used {\sc photutils} \citep{bradley22}, an {\sc astropy}-affiliated \citep{astropy13,astropy18} {\sc python} package that provides tools for performing photometry. 

\begin{figure*}
\begin{center}
\includegraphics[width=15cm,angle=0]{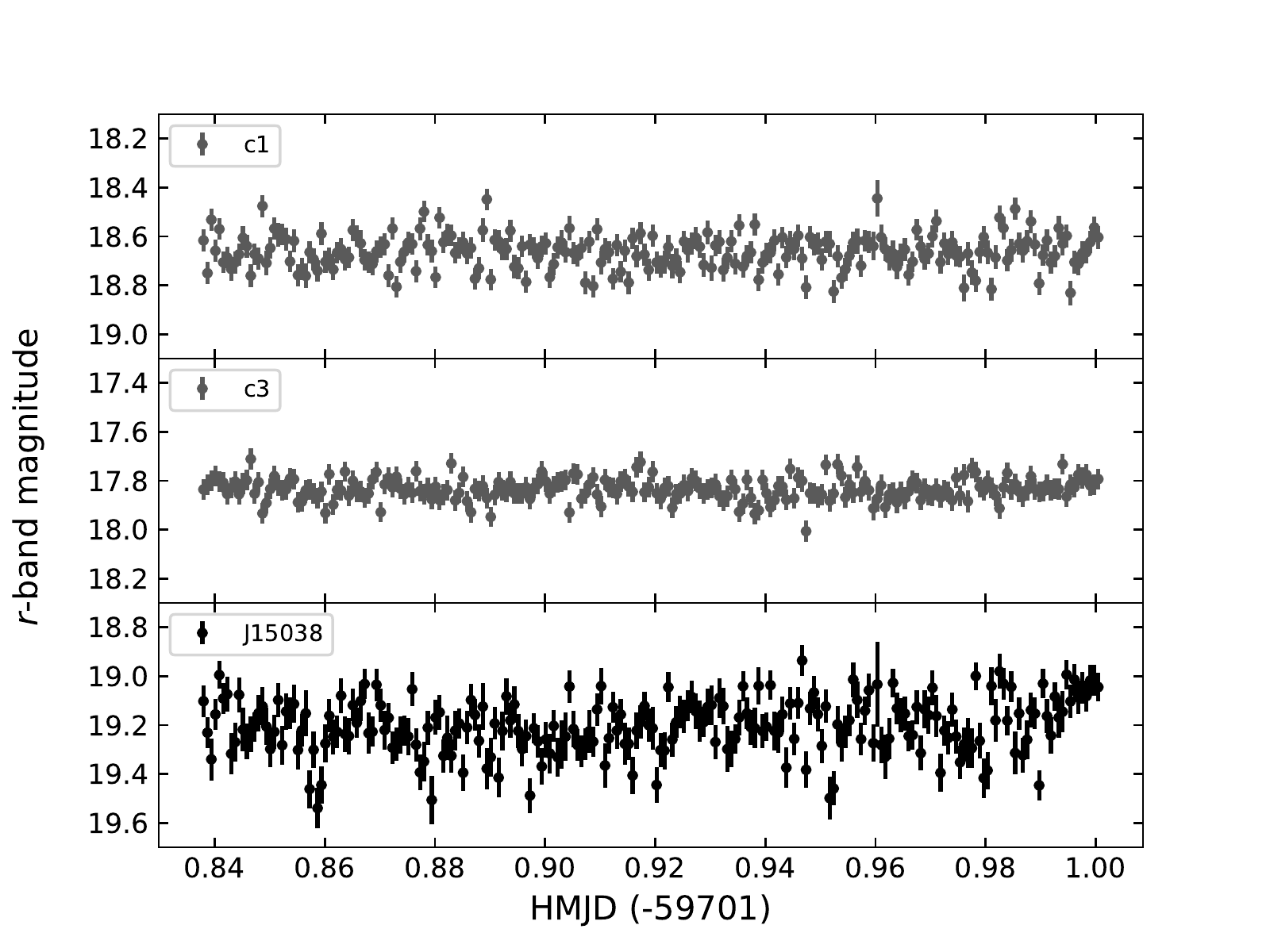}
\caption{Lower panel: LCO Sinistro $r'$-band light curve of the optical counterpart to J15038. The upper and middle panels show the $r'$-band light curves of two of the comparison stars used to calibrate the photometry of the target.\label{fig:lco}}
\end{center}
\end{figure*}

For each image we built an effective Point Spread Function \citep[ePSF;][]{anderson00} by selecting $\sim130$ bright, isolated stars. We iterated between the ePSF and the stars used to build it 10 times, oversampling with respect to the detector pixels by a factor of 4. For each image, the ePSF was then used to fit and subtract the detected sources, resulting in a best-fit flux for each detected source. To calibrate the photometry of the target, we selected eight field stars as comparison sources and used their known $r'$-band magnitudes from the American Association of Variable Star Observers (AAVSO) Photometric All-Sky Survey \citep[APASS;][]{henden16} to derive the magnitude of the target. The light curve of the optical counterpart to J15038 is shown in Figure~\ref{fig:lco}, along with the light curves of two of the comparison stars.

\section{Results}

\subsection{X-ray Timing}

\begin{figure*}
\begin{center}
\includegraphics[width=15cm]{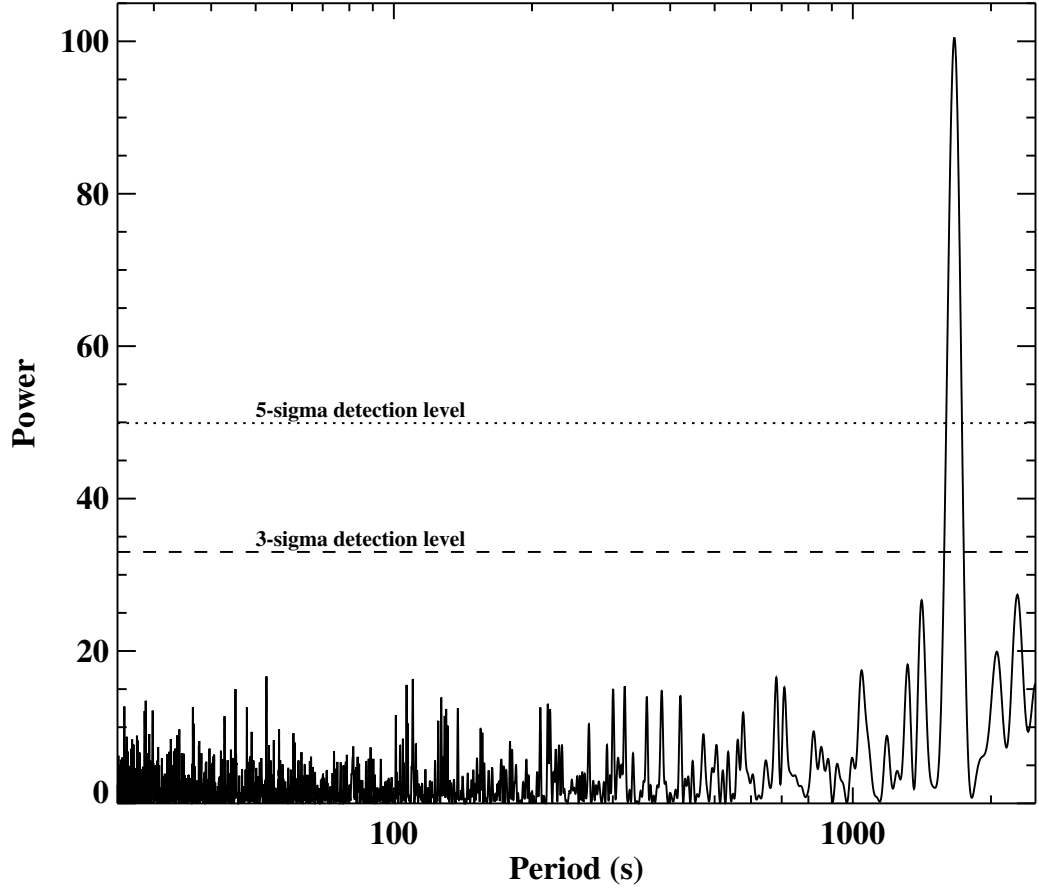}
\caption{Periodogram of J15038 for 0.3-12\,keV {\em XMM} (pn, MOS1, and MOS2) calculated using the $Z_{1}^{2}$ test.  A period of $1664\pm 8$\,s is detected at high significance.\label{fig:periodogram}}
\end{center}
\end{figure*}

As many magnetic CVs show periodicities in their light curves at the spin period of the WD, we carried out a periodicity search.  We made an event list for the {\em XMM} instruments (pn, MOS1, and MOS2) composed of photons in the source regions in the 0.3-12\,keV bandpass, and we used {\ttfamily barycen} to shift the individual event times to the Solar system barycenter.  Using the $Z_{1}^{2}$ (Rayleigh) test \citep{buccheri83}, we calculated the power for 40,000 frequencies ranging from $4\times 10^{-4}$ to 0.04\,Hz.  The 25-2500\,s periodogram is shown in Figure~\ref{fig:periodogram}, and a strong signal at $1664\pm 8$\,s (1-$\sigma$ uncertainty) is detected.  The peak of the signal is at a power of $S=100.5$, and the false alarm probability (FAP) is given by $e^{-S/2}$ multiplied by the number of trials.  In this case, we calculate a FAP of $6\times 10^{-18}$, indicating a highly significant signal.  The errors on the period are calculated by finding the locations where the periodogram has values of $S$--1 (above and below 1664\,s).  


We repeated the analysis with data from the two {\em NuSTAR} instruments (FPMA and FPMB) by making an event list composed of the arrival times (shifted to the Solar system barycenter) of photons detected in the source regions in the 3-79\,keV bandpass.  In addition to strong peaks at harmonics of the {\em NuSTAR} orbital period due to Earth occultations of J15038, the 25-2500\,s periodogram also shows a peak near the period found in the {\em XMM} data.  The peak is at $1678\pm 2$\,s (1$\sigma$ uncertainty) with a power of $S=76.5$. If we count trials using the full number of frequencies in the periodogram (40,000), $S=76.5$ corresponds to a FAP of $9\times 10^{-13}$.  However, if we consider that we were searching for a period near the {\em XMM} period, the true number of trials is closer to 10, which leads to a FAP of $2\times 10^{-16}$.

\begin{figure*}
\begin{center}
\includegraphics[width=15cm]{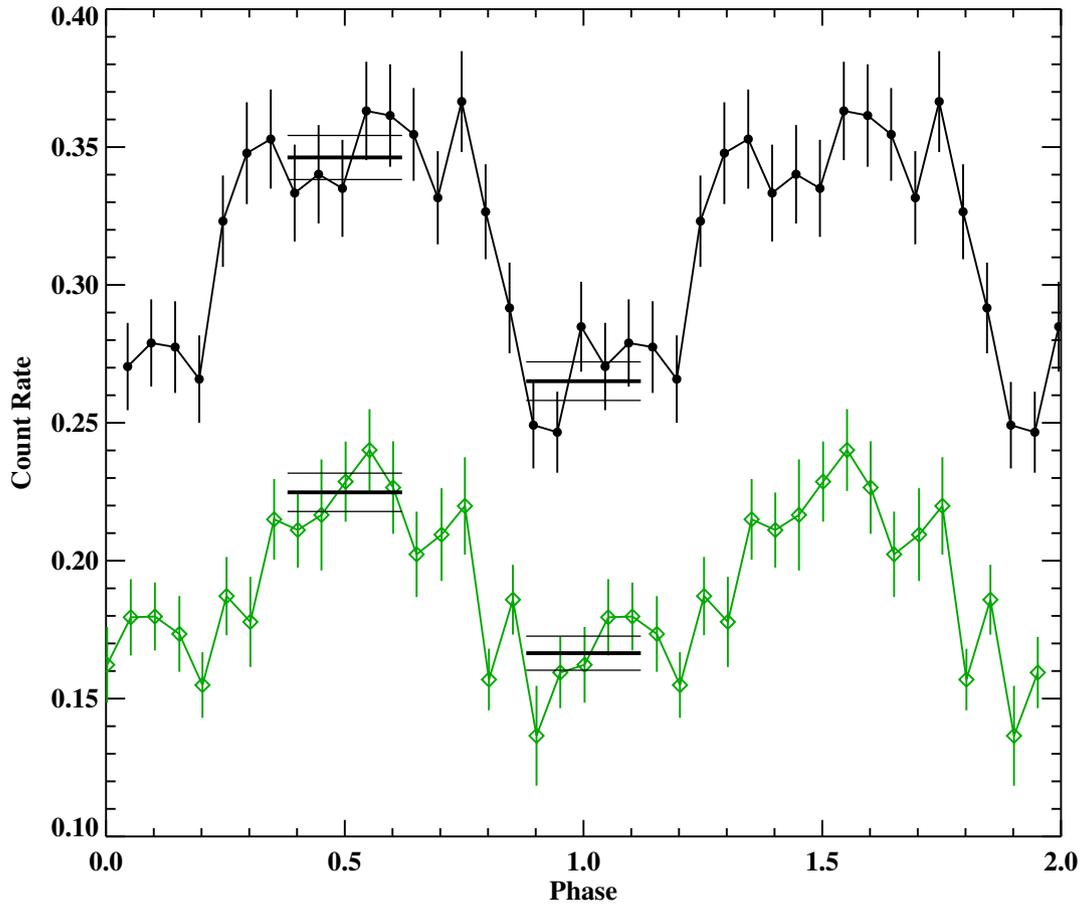}
\caption{\small Light curves folded on a period of 1678\,s.  The 0.3-12\,keV {\em XMM} (pn, MOS1, and MOS2 added) data are shown in black with filled circles.  The 3-12\,keV {\em NuSTAR} data are shown in green with diamonds.  The thick solid horizontal lines show the averages of the five points closest to phase zero (=1) and phase 0.5.  The thin solid horizontal lines show the 1$\sigma$ uncertainties on the averages.  The plotted rates are after the background rates are subtracted.  The same phase zero time is used for {\em XMM} and {\em NuSTAR}.\label{fig:folded_energy}}
\end{center}
\end{figure*}
\begin{figure*}
\begin{center}
\includegraphics[width=15cm]{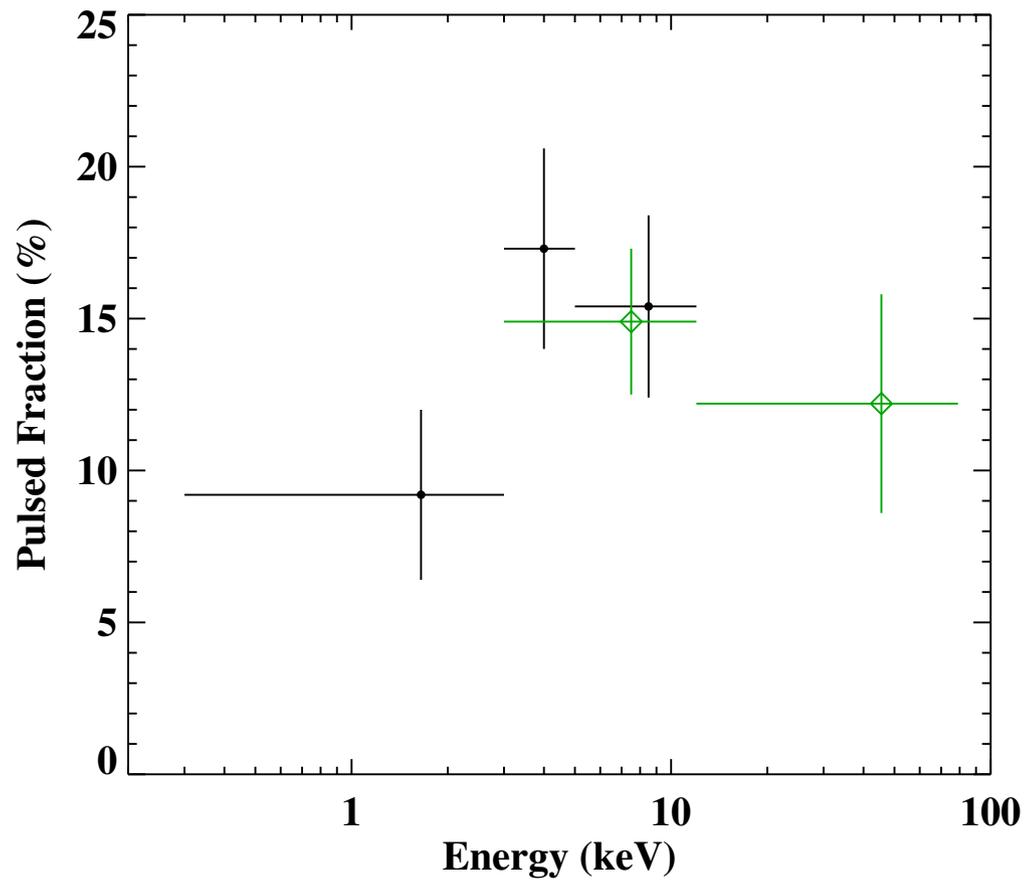}
\caption{\small The fractional amplitudes of the 1678\,s periodic signal measured with {\em XMM} (black filled circles) and with {\em NuSTAR} (green diamonds). The error bars are at the 1-$\sigma$ confidence level.\label{fig:amplitude}}
\end{center}
\end{figure*}

Due to the longer duration of the {\em NuSTAR} observation compared to the {\em XMM} observation, the period measured by {\em NuSTAR} is more precise, and we take $1678\pm 2$\,s to be the best estimate of the period of the signal.  To determine the amplitude of the signal, we folded the {\em XMM} (pn, MOS1, and MOS2 added) and {\em NuSTAR} (FPMA and FPMB added) light curves on the 1678\,s period.  The amplitude is determined by calculating the average background subtracted count rate of five bins closest to phase zero ($C_{\rm min}$) and the average count rate of the five bins closest to phase 0.5 ($C_{\rm max}$).  For {\em XMM} in the full 0.3-12\,keV bandpass, the amplitude is ($C_{\rm max}-C_{\rm min}$)/($C_{\rm max}+C_{\rm min}$) = 13.3\% $\pm$ 1.7\% (1$\sigma$ confidence error), and the amplitude is 16.2\% $\pm$ 2.2\% at 3-12\,keV.  For {\em NuSTAR}, the amplitude is 14.9\% $\pm$ 2.4\% at 3-12\,keV, showing good agreement between {\em XMM} and {\em NuSTAR}.  The {\em XMM} and {\em NuSTAR} folded light curves shown in Figure~\ref{fig:folded_energy} have both been folded using the same phase zero ephemeris of MJD 59060.49504.  We repeated this calculation for three {\em XMM} and two {\em NuSTAR} energy bands, and Figure ~\ref{fig:amplitude} shows the fractional pulse amplitude vs. energy.  


\subsection{X-ray Spectrum}

For the spectral analysis, we used the XSPEC v12.12.0 package \citep{arnaud96} and performed the fitting with $\chi^2$ minimization.   We binned the spectra by requiring bins with detection significance greater than 5$\sigma$ for pn, FPMA, and FPMB and greater than 3$\sigma$ for MOS1 and MOS2.  We first fitted the {\em XMM} (pn, MOS1, and MOS2) and {\em NuSTAR} (FPMA and FPMB) spectra with an absorbed power-law model, using \cite{wam00} abundances. We also allowed for normalization differences between instruments by introducing a multiplicative constant, making the overall model \begin{ttfamily}constant*tbabs*pegpwrlw\end{ttfamily} in XSPEC notation. The resulting fit shows that the spectrum is hard with $\Gamma \sim 1.1$. However, the fit is poor with $\chi^2$ = 1333 for 582 degrees of freedom (dof), with large residuals appearing in the iron line region (6-7\,keV), indicating a strong emission line or lines.  Large residuals are also present at the lowest (below 1\,keV) and highest (above 30\,keV) energies.  Adding a broad Gaussian significantly improves the fit to $\chi^2/$dof = 833/579 (see Table~\ref{tab:quality}). However, the lower $\chi^{2}$ is due to the Gaussian improving the fit to the continuum rather than properly fitting the iron line.  Also, there are negative residuals in the high energy part of the spectrum, which indicates curvature in the spectrum. 

To allow for curvature in the model and to incorporate the emission mechanism thought to operate in IPs, we changed the continuum model to a bremsstrahlung component. While the model \begin{ttfamily}constant*tbabs*(gaussian+bremss)\end{ttfamily} does not provide a good fit to the data $(\chi^2/$dof = 1020/579), adding a partial covering absorption component to the model \citep{srr05} (\begin{ttfamily}pcfabs\end{ttfamily}) with $N_{\rm H} = (8.2^{+1.9}_{-1.1})\times 10^{22}$ cm$^{-2}$ and a covering fraction of $0.69^{+0.03}_{-0.02}$ (90\% confidence uncertainties are quoted unless otherwise indicated) improves the fit greatly to $\chi^2/$dof = 568/577. With this model, the bremsstrahlung temperature is very high, $kT > 117$\,keV (90\% confidence limit), but there are still residuals at the high energy end that we suspect are related to the presence of a reflection component.

Reflection of the hard X-ray emission off the WD surface is often included in models when fitting IP spectra, and {\em XMM} and {\em NuSTAR} spectra show strong evidence for this component in other IPs \citep{mukai15,tomsick16b,shaw18}. To incorporate this into our working model, we added a reflection component using the \begin{ttfamily}reflect\end{ttfamily} model in XSPEC. This component is based on \cite{mz95}, which is for reflection of direct emission from neutral material and includes dependence on inclination. By convolving \begin{ttfamily}bremss\end{ttfamily} with \begin{ttfamily}reflect\end{ttfamily}, the model includes both a direct and a reflected bremsstrahlung component where the strength of the reflected component depends on the amplitude parameter, $\Omega/2\pi$. In our case, if the reflection amplitude is left as a free parameter, it will increase to values above 1.0. However, assuming that we see 100\% of the direct emission, values of $\Omega/2\pi$ above unity are not physically possible for reflection from the WD surface. Thus, we fix the reflection amplitude to 1.0, resulting in a fit with $\chi^2/$dof=541/575. Although the bremsstrahlung temperature drops
to $kT = 57^{+39}_{-18}$\,keV when we add \begin{ttfamily}reflect\end{ttfamily}, it is still high for an IP.

While the actual reflection component has three main features: the Compton hump above 10\,keV, the iron edge at 7.1\,keV (for neutral iron), and an iron fluorescence line at 6.4\,keV (also for neutral iron), the \cite{mz95} model only includes the first two features, which is one reason that we include the Gaussian in the model above. If the emission line was due only to reflection from the WD surface, we would expect a relatively narrow line centered at 6.4\,keV; however, as shown in Table~\ref{tab:spec1}, we measure a broad line with $\sigma_{\rm line}=0.32^{+0.06}_{-0.05}$\,keV at an energy above 6.4\,keV ($E_{\rm line} = 6.52 \pm 0.05$\,keV). It is very likely that this is due to contributions from higher ionization states coming from the hot material in the accretion column, and this is common for IPs \citep{hm04}.  We modified the model to include three lines with energies fixed to 6.4\,keV (neutral iron) and 6.7\,keV (He-like iron), and 6.97\,keV (H-like iron). We fixed the line widths to 50\,eV, which is, in effect, a narrow line since it is much smaller than the {\em XMM} and {\em NuSTAR} energy resolutions.  This more physical representation of the emission lines provides an equally good fit $\chi^2/$dof = 542/575.  The equivalent widths of the three lines are $273\pm 44$, $109\pm 35$, and $104\pm 43$\,eV for neutral, He-like, and H-like iron, respectively.

We replaced the bremsstrahlung component with a physically-motivated IP model with WD mass ($M_{\rm WD}$) as a free parameter.  A constraint on $M_{\rm WD}$ is possible because $M_{\rm WD}$ sets the maximum temperature in the post-shock regions (PSRs) of the accretion column.  While we used the IP mass (IPM) model \citep{srr05} in earlier work \citep{tomsick16b}, this model has been updated to include a second free parameter, $R_{\rm m}$, which corresponds to the radius where the accretion disk is disrupted by the WD magnetic field.  The value of $R_{\rm m}$ affects the calculation of the maximum velocity of the material in the accretion column, which can in turn affect the value of $M_{\rm WD}$. However, the effect is only significant if $R_{\rm m}$ is a few WD radii ($R_{\rm WD}$) or less.  For $R_{\rm m}\gsim 10$\,$R_{\rm WD}$, the effect on $M_{\rm WD}$ is less than a few percent \citep{shaw20}.  A detailed description of this model is found in \cite{suleimanov16}, where it is referred to as the PSR model, and the PSR model was also used extensively in \cite{shaw20}.  For J15038, the results for a model including direct and reflected PSR components, partial covering, and three Gaussians are reported in Table~\ref{tab:spec2}. The value of $R_{\rm m} = 107\,R_{\rm WD}$ is calculated assuming that $R_{\rm m}$ is equal to the co-rotation radius (see equations 3 and 4 in \citealt{suleimanov16}).  The calculation uses the WD spin period of 1678\,s and the $M_{\rm WD}$ parameter value.  The PSR model parameters indicate a high WD mass, $M_{\rm WD} = 1.36^{+0.04}_{-0.11}$\Msun~(90\% confidence errors), which is in agreement with the high bremsstrahlung temperature found in the earlier fits.  


\subsection{Optical Light Curve}

\begin{figure*}
\begin{center}
    \includegraphics[width=15cm]{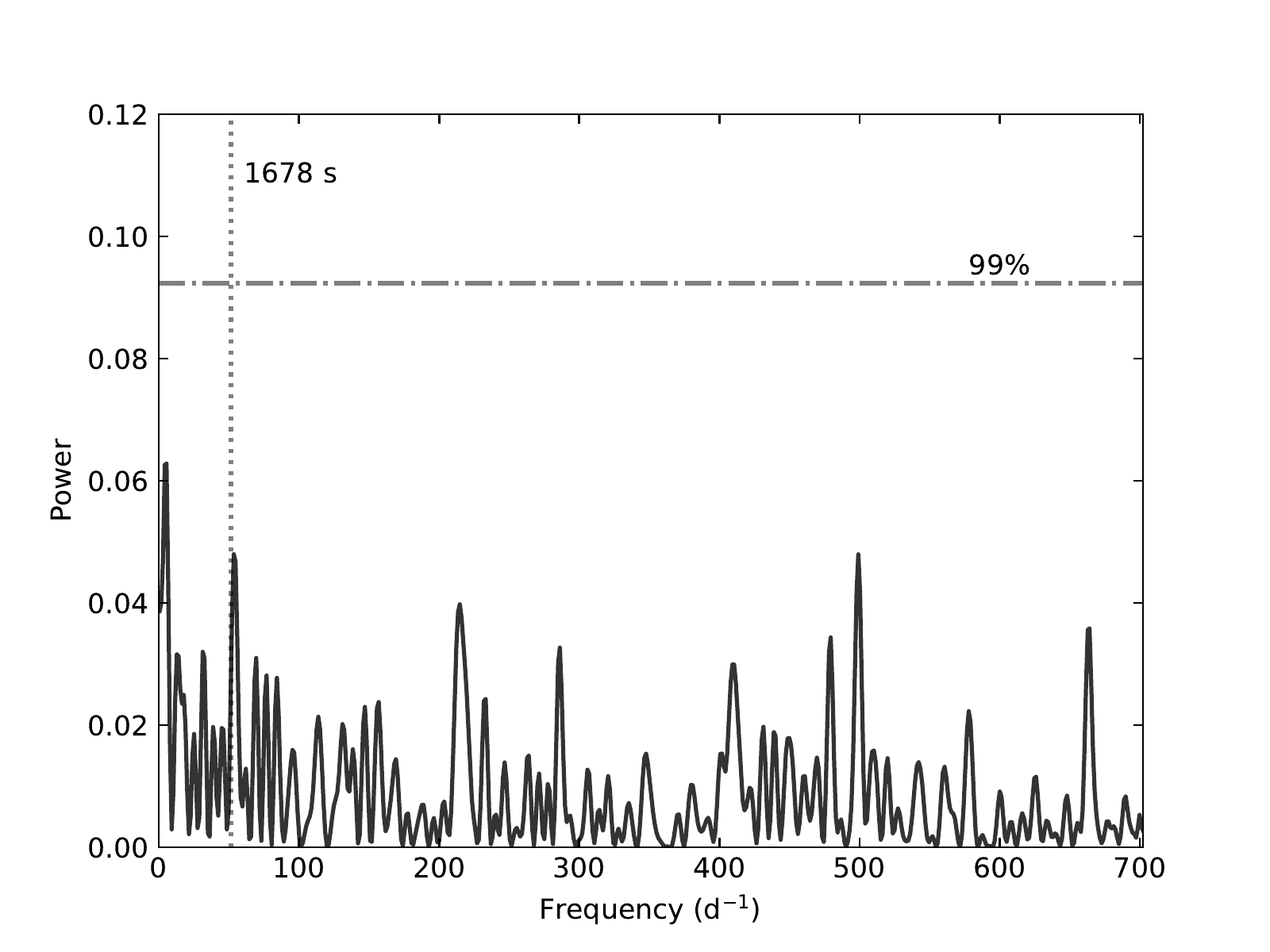}
    \caption{Lomb Scargle periodogram of the optical counterpart to J15038, with the X-ray periodicity, likely the spin period of the WD, marked by the vertical dotted line. The horizontal dot-dashed line represents the 99\% significance threshold for the power.\label{fig:LCO_periodograms}}
\end{center}
\end{figure*}

The $r'$-band light curve of the optical counterpart to J15038 is shown in Figure \ref{fig:lco}, along with the light curves of two of the eight comparison stars we used to calibrate the photometry of the target. With exposure time and overheads, the light curves have a cadence of $\approx1$ minute. 
We see that J15038 potentially shows some intrinsic variability when compared to the comparison stars. We search for periodic variability by calculating the Lomb-Scargle periodogram \citep{lomb76,scargle82}, which utilizes least-squares fitting of sinusoids to the light curve to determine the power at each frequency in a given range. We implemented this using the {\tt LombScargle} class in {\sc astropy}. The periodogram is shown in Figure~\ref{fig:LCO_periodograms}. We calculated the 99\% significance threshold for the power by randomly shuffling the light-curve magnitudes but keeping the time stamps the same for each epoch, effectively creating a randomized light curve with the same sampling as the original data. We calculated the peak Lomb–Scargle power for 10,000 of these randomized light curves, from which we derived the 99\% significance level and plotted it in Figure~\ref{fig:LCO_periodograms}. We find that no peaks appear above this line and thus we conclude that we are unable to detect a periodicity in the $r'$-band light curve of J15038. 


\subsection{Update on Optical and Near-IR Photometry and the Source Distance}

In \cite{tomsick20}, we used an upper limit on $N_{\rm H}$ and near-IR magnitudes to derive a range of effective temperatures (assuming thermal near-IR emission) between 4000 and 6500\,K.  With the improvement in the $N_{\rm H}$ measurement provided by the {\em XMM}+{\em NuSTAR} spectrum to $(5.0^{+0.7}_{-0.6})\times 10^{21}$\,cm$^{-2}$, the extinction is much better constrained to $A_{V} = 2.3\pm 0.3$ (using the \cite{go09} conversion factor), allowing us to re-derive the temperature using optical magnitudes from {\em Gaia} DR3 \citep{gaia_edr3,gaia_dr3}: $BP=20.038\pm 0.040$ and $RP = 17.762\pm 0.021$.  Using these magnitudes and the relation $A_{BP}-A_{RP} = 0.51 A_{V}$ \citep{fitzpatrick99}, we find that the extinction corrected color is $BP-RP-(A_{BP}-A_{RP}) = 1.11\pm 0.16$.  Using the color/temperature tables from \cite{pm13} gives a temperature between 4700 and 5400\,K.  This significantly improves the constraint on the temperature.  While it is not clear if this temperature corresponds to that of the accretion disk or the companion star, if it is the accretion disk, then the star must be cooler.

{\em Gaia} DR3 also provides a revised distance to the source of $1.3^{+0.7}_{-0.4}$\,kpc \citep{bj21}, which is only slightly larger than the {\em Gaia} DR2 distance of $1.1^{+1.5}_{-0.4}$\,kpc.  In both cases, these are the geometric distances since the colors of J15038 may deviate from the assumptions made for the photogeometric distances.  Using the revised distance, $A_{G} = 2.0\pm 0.3$, and $G = 19.036\pm 0.005$ gives an absolute magnitude of $M_{G} = 6.4^{+0.7}_{-1.1}$.  This indicates that the spectral type of the companion is later than G9V \citep{pm13}. 

\section{Discussion and Conclusions}

Our analysis of the {\em XMM} and {\em NuSTAR} data from an observation of J15038 confirms the previously suspected IP nature.  We have uncovered a $1678\pm 2$\,s periodicity that we associate with the WD spin period.  The signal is strongest in the 3-12\,keV band with an amplitude near 17\%.  It is weaker below 3\,keV and above 12\,keV, but it is detected across the {\em XMM} and {\em NuSTAR} energy bands.  The relatively weak dependence of the signal strength with energy is somewhat unusual for IPs.  More often, IP signal amplitudes are highest at low energies as has been seen for AO~Psc, V1223~Sgr, and J14091 \citep{taylor97,hayashi11,tomsick16b}.  

The {\em XMM} and {\em NuSTAR} energy spectrum also provides the first detection of iron emission lines from J15038.  Although multiple lines are not resolved, it is clear that multiple lines are required, and the spectrum is well-described by a strong line at 6.4\,keV with an equivalent width of $273\pm 4$\,eV and lines at 6.7\,keV and 6.97\,keV with equivalent widths near 100\,eV.  We interpret the 6.4\,keV line as being due to reflection of the hard X-ray emission from the accretion column off the WD surface, while the higher-energy lines come from He-like (6.7\,keV) and H-like (6.97\,keV) iron in the accretion column.  The ratio of the fluxes of these two lines provides another diagnostic of the plasma temperature and potentially $M_{\rm WD}$ \citep{xu19}.  For J15038, the results in Tables~\ref{tab:spec1} and \ref{tab:spec2} give a ratio of the 6.97\,keV to 6.7\,keV line of $0.7\pm 0.4$ (90\% confidence error).  Based on results from other IPs presented in \cite{xu19}, the corresponding range of WD masses would be from 0.4\Msun~to $>$1.1\Msun.  Unfortunately, the errors are too large to provide a meaningful constraint since the line flux ratio of $0.7\pm 0.4$ essentially spans the entire ratio range for the sample of IPs studied by \cite{xu19}.  It is explained in \cite{yu22} that the iron line ratio method for determining WD masses has limitations at the high-mass end due to the thermal plasma model used and the fact that the most massive WD in their sample has a mass of 1.2\Msun.

\begin{figure*}
\begin{center}
\includegraphics[width=15cm]{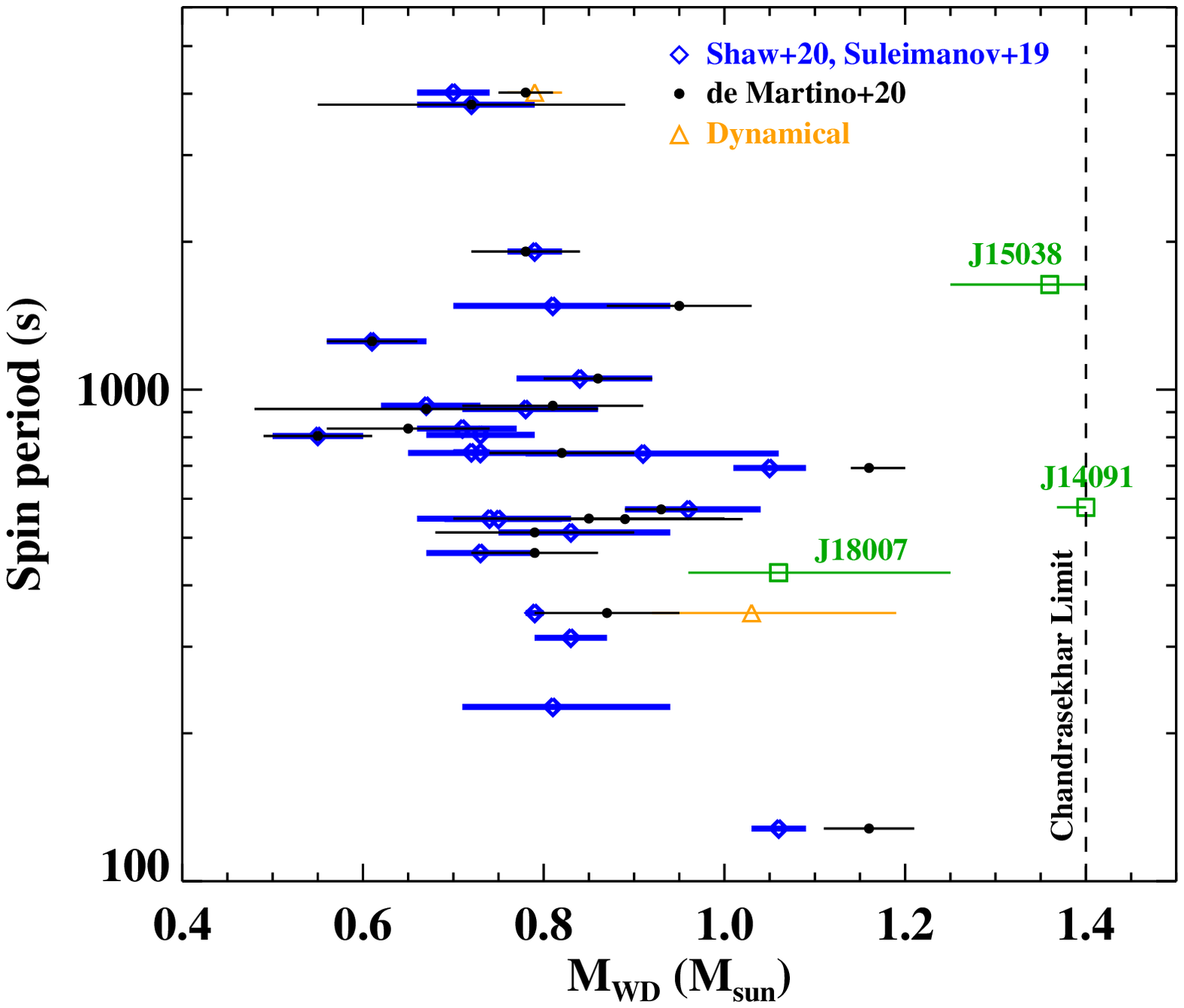}
    \caption{The WD masses and spin periods of IPs.  The sources marked with blue diamonds have WD masses from \protect\cite{shaw20} or \protect\cite{suleimanov19} and periods from \protect\cite{rk11}.  The sources marked with black filled circles are from \protect\cite{demartino20}.  The sources in green are from our studies of IGR sources:  J14091 \protect\citep{tomsick16b}; J18007 \protect\citep{coughenour22}; and J15038 (this work).  The references for the dynamical mass measurements are provided in the text.\label{fig:spin_and_mass}}
\end{center}
\end{figure*}

The energy spectrum that we fit in this work has also allowed for the details of the continuum to become clear.  We measure a high bremsstrahlung temperature for the material in the accretion column and also use the PSR model to constrain $M_{\rm WD}$. These fits indicate that $M_{\rm WD}$ is at least 1.25\Msun~(90\% confidence lower limit) and is also consistent with approaching the Chandrasekhar limit ($M_{\rm WD} = 1.36^{+0.04}_{-0.11}$\Msun).  The comparison sources shown in Figure~\ref{fig:spin_and_mass} are the IPs\footnote{Based on the classification as an IP or and IP candidate at https://asd.gsfc.nasa.gov/Koji.Mukai/iphome/iphome.html} with WD mass measurements obtained from {\em NuSTAR} observations.  \cite{shaw20} and \cite{suleimanov19} find values of $M_{\rm WD}$ between 0.55\Msun~and 1.06\Msun~and an average of 0.77\Msun~\citep{shaw20}.  While this average mass is already higher than the average masses of isolated WDs and pre-CVs, suggesting that WDs gain mass during the accretion-outburst-accretion cycle, J15038 has a much higher mass.  In addition, the mass of J14091 from \cite{tomsick16b} is another IGR IP with much higher mass than the previously observed IP population.  The third source that has been observed as part of this program, J18007, has properties that fall within the distribution of other WD masses in IPs, but, at $1.06^{+0.19}_{-0.10}$\Msun, it is at the higher end \citep{coughenour22}.

Figure~\ref{fig:spin_and_mass} also shows that the J15038 spin period of 1678\,s is longer than average but not atypical with three spin periods being longer (TV~Col at 1910\,s, V1062~Tau at 3800\,s, and EX~Hya at 4022\,s).  Showing the spin periods and WD masses on the same plot provides a convenient way to compare both quantities to other IPs.  We do not necessarily expect to see a relationship between spin and mass, and the plot is not meant to suggest that there is any relationship.

In addition to the {\em NuSTAR}-measured masses \citep{suleimanov19,shaw20}, Figure~\ref{fig:spin_and_mass} also shows WD masses listed in \cite{demartino20} as well as dynamical WD masses.  The \cite{demartino20} masses are measured by modeling the X-ray spectra, but they are independent of the {\em NuSTAR}-measured masses since they come from modeling spectra from other satellites.  Two of the IPs shown in Figure~\ref{fig:spin_and_mass} have dynamical WD mass measurements: EX~Hya \citep{br08}; and GK~Per \citep{alvarez21}.  In some cases, the \cite{demartino20} and dynamical masses are somewhat higher than the {\em NuSTAR}-measured values.  As more dynamical masses become available \citep[e.g.,][]{alvarez23}, it will be interesting to make additional comparisons.  However, based on the data shown in Figure~\ref{fig:spin_and_mass}, it is clear that J15038 and J14091 are outliers when compared to the general population.  

In the standard binary evolution scenario, J15038 and J14091 either were born with high WD masses, or they have undergone significant mass growth via accretion.  The topic of high WD birth masses is related to the maximum mass of progenitor stars that result in WDs, and evidence has been found that WDs can have initial masses of at least 1.2\Msun~\citep{miller22}.  This is still lower than the masses we are finding for J15038 and J14091, suggesting that even if they are born with large masses, it is likely that, over time, they gain more mass by accretion than they lose during nova eruptions.  

Concerning the question of the progenitors of type Ia SNe, seeing massive WDs in accreting binaries (J15038 and J14091) is a point in favor of the SD accretion channel for type Ia SNe.  However, the number of known isolated massive WDs is also increasing with discoveries of sources like ZTF~J190132.9+145808.7 \citep{caiazzo21} and the rapidly rotating (70\,s spin period) SDSS~J221141.80+113604.4 \citep{kilic21}, which both are likely to be the product of past WD mergers (the DD channel).  SD and DD channels are both still viable, and continuing to improve our knowledge of the accreting and isolated WD mass distributions is an important part of determining which is the dominant channel.

Our results on the WD masses for J15038 and J14091 suggest that there may be a larger population of high-mass WDs among the IGR sources.  While an {\em INTEGRAL} detection for a IP indicates a relatively high temperature in the accretion column, it is only with the increase in hard X-ray sensitivity provided by {\em NuSTAR} that allows for spectra with high enough quality to constrain the bremsstrahlung temperatures and therefore the WD masses.  There are still a large number of unclassified IGR sources \citep{krivonos22}, and the number of IGR IPs will increase.  More work is necessary to find additional IPs, to confirm their IP nature with WD spin measurements with {\em XMM}, and to measure their masses with {\em NuSTAR}. The end result may be a WD mass distribution with a significantly larger number of high-mass WDs than previous studies have found.

\section*{Acknowledgments}

JAT acknowledges partial support from NASA under {\em NuSTAR} Guest Observer grant 80NSSC21K0064.  BMC acknowledges partial support under NASA contract NNG08FD60C.  JH acknowledges support from an appointment to the NASA Postdoctoral Program at the Goddard Space Flight Center, administered by the ORAU through a contract with NASA.  MC acknowledges financial support from the Centre National d’Etudes Spatiales (CNES). RK acknowledges support from the Russian Science Foundation (grant 19-12-00396). This work made use of data from the {\em NuSTAR} mission, a project led by the California Institute of Technology, managed by the Jet Propulsion Laboratory, and funded by the National Aeronautics and Space Administration. This research has made use of the {\em NuSTAR} Data Analysis Software (NuSTARDAS) jointly developed by the ASI Science Data Center (ASDC, Italy) and the California Institute of Technology (USA). This research made use of Photutils, an Astropy package for detection and photometry of astronomical sources \citep{bradley22}. This research made use of Astropy,\footnote{\href{http://www.astropy.org}{http://www.astropy.org}} a community-developed core Python package for Astronomy \citep{astropy13, astropy18}.

\section*{Data Availability}

Data used in this paper are available through NASA's HEASARC and the {\em XMM-Newton} Science Archive (XSA).



\begin{thebibliography}{74}
\expandafter\ifx\csname natexlab\endcsname\relax\def\natexlab#1{#1}\fi

\bibitem[{{{\'A}lvarez-Hern{\'a}ndez}
  {et~al}\mbox{.}(2021){{\'A}lvarez-Hern{\'a}ndez}, {Torres},
  {Rodr{\'\i}guez-Gil}, {Shahbaz}, {Anupama}, {Gazeas}, {Pavana}, {Raj},
  {Hakala}, {Stone}, {Gomez}, {Jonker}, {Ren}, {Cannizzaro},
  {Pastor-Marazuela}, {Goff}, {Corral-Santana}, \& {Sabo}}]{alvarez21}
{{\'A}lvarez-Hern{\'a}ndez} A. {et~al.}, 2021, MNRAS, 507, 5805

\bibitem[{{{\'A}lvarez-Hern{\'a}ndez}
  {et~al}\mbox{.}(2023){{\'A}lvarez-Hern{\'a}ndez}, {Torres},
  {Rodr{\'\i}guez-Gil}, {Shahbaz}, {S{\'a}nchez-Sierras}, {Acosta-Pulido},
  {Jonker}, {Gazeas}, {Hakala}, \& {Corral-Santana}}]{alvarez23}
{{\'A}lvarez-Hern{\'a}ndez} A. {et~al.}, 2023, arXiv:2304.08524

\bibitem[{{Anderson} \& {King}(2000)}]{anderson00}
{Anderson} J., {King} I.~R., 2000, PASP, 112, 1360

\bibitem[{{Arnaud}(1996)}]{arnaud96}
{Arnaud} K.~A., 1996, in Astronomical Society of the Pacific Conference Series,
  Vol. 101, Astronomical Data Analysis Software and Systems V, {Jacoby} G.~H.,
  {Barnes} J., eds., p.~17

\bibitem[{{Astropy Collaboration} {et~al}\mbox{.}(2018){Astropy Collaboration},
  {Price-Whelan}, {Sip{\H o}cz}, {G{\"u}nther}, {Lim}, {Crawford}, {Conseil},
  {Shupe}, {Craig}, {Dencheva}, {Ginsburg}, {VanderPlas}, {Bradley},
  {P{\'e}rez-Su{\'a}rez}, {de Val-Borro}, {Aldcroft}, {Cruz}, {Robitaille},
  {Tollerud}, {Ardelean}, {Babej}, {Bach}, {Bachetti}, {Bakanov}, {Bamford},
  {Barentsen}, {Barmby}, {Baumbach}, {Berry}, {Biscani}, {Boquien}, {Bostroem},
  {Bouma}, {Brammer}, {Bray}, {Breytenbach}, {Buddelmeijer}, {Burke},
  {Calderone}, {Cano Rodr{\'{\i}}guez}, {Cara}, {Cardoso}, {Cheedella},
  {Copin}, {Corrales}, {Crichton}, {D'Avella}, {Deil}, {Depagne}, {Dietrich},
  {Donath}, {Droettboom}, {Earl}, {Erben}, {Fabbro}, {Ferreira}, {Finethy},
  {Fox}, {Garrison}, {Gibbons}, {Goldstein}, {Gommers}, {Greco}, {Greenfield},
  {Groener}, {Grollier}, {Hagen}, {Hirst}, {Homeier}, {Horton}, {Hosseinzadeh},
  {Hu}, {Hunkeler}, {Ivezi{\'c}}, {Jain}, {Jenness}, {Kanarek}, {Kendrew},
  {Kern}, {Kerzendorf}, {Khvalko}, {King}, {Kirkby}, {Kulkarni}, {Kumar},
  {Lee}, {Lenz}, {Littlefair}, {Ma}, {Macleod}, {Mastropietro}, {McCully},
  {Montagnac}, {Morris}, {Mueller}, {Mumford}, {Muna}, {Murphy}, {Nelson},
  {Nguyen}, {Ninan}, {N{\"o}the}, {Ogaz}, {Oh}, {Parejko}, {Parley}, {Pascual},
  {Patil}, {Patil}, {Plunkett}, {Prochaska}, {Rastogi}, {Reddy Janga},
  {Sabater}, {Sakurikar}, {Seifert}, {Sherbert}, {Sherwood-Taylor}, {Shih},
  {Sick}, {Silbiger}, {Singanamalla}, {Singer}, {Sladen}, {Sooley},
  {Sornarajah}, {Streicher}, {Teuben}, {Thomas}, {Tremblay}, {Turner},
  {Terr{\'o}n}, {van Kerkwijk}, {de la Vega}, {Watkins}, {Weaver}, {Whitmore},
  {Woillez}, {Zabalza}, \& {Astropy Contributors}}]{astropy18}
{Astropy Collaboration} {et~al.}, 2018, AJ, 156, 123

\bibitem[{{Astropy Collaboration} {et~al}\mbox{.}(2013){Astropy Collaboration},
  {Robitaille}, {Tollerud}, {Greenfield}, {Droettboom}, {Bray}, {Aldcroft},
  {Davis}, {Ginsburg}, {Price-Whelan}, {Kerzendorf}, {Conley}, {Crighton},
  {Barbary}, {Muna}, {Ferguson}, {Grollier}, {Parikh}, {Nair}, {Unther},
  {Deil}, {Woillez}, {Conseil}, {Kramer}, {Turner}, {Singer}, {Fox}, {Weaver},
  {Zabalza}, {Edwards}, {Azalee Bostroem}, {Burke}, {Casey}, {Crawford},
  {Dencheva}, {Ely}, {Jenness}, {Labrie}, {Lim}, {Pierfederici}, {Pontzen},
  {Ptak}, {Refsdal}, {Servillat}, \& {Streicher}}]{astropy13}
{Astropy Collaboration} {et~al.}, 2013, A\&A, 558, A33

\bibitem[{{Bailer-Jones} {et~al}\mbox{.}(2021){Bailer-Jones}, {Rybizki},
  {Fouesneau}, {Demleitner}, \& {Andrae}}]{bj21}
{Bailer-Jones} C.~A.~L., {Rybizki} J., {Fouesneau} M., {Demleitner} M.,
  {Andrae} R., 2021, AJ, 161, 147

\bibitem[{{Beuermann} \& {Reinsch}(2008)}]{br08}
{Beuermann} K., {Reinsch} K., 2008, A\&A, 480, 199

\bibitem[{{Bird} {et~al}\mbox{.}(2016){Bird}, {Bazzano}, {Malizia}, {Fiocchi},
  {Sguera}, {Bassani}, {Hill}, {Ubertini}, \& {Winkler}}]{bird16}
{Bird} A.~J. {et~al.}, 2016, ApJS, 223, 15

\bibitem[{{Bradley} {et~al}\mbox{.}(2022){Bradley}, {Sip{\H{o}}cz},
  {Robitaille}, {Tollerud}, {Vin{\'\i}cius}, {Deil}, {Barbary}, {Wilson},
  {Busko}, {Donath}, {G{\"u}nther}, {Cara}, {Lim}, {Me{\ss}linger}, {Conseil},
  {Bostroem}, {Droettboom}, {Bray}, {Andersen Bratholm}, {Barentsen}, {Craig},
  {Rathi}, {Pascual}, {Perren}, {Georgiev}, {De Val-Borro}, {Kerzendorf},
  {Bach}, {Quint}, \& {Souchereau}}]{bradley22}
{Bradley} L. {et~al.}, 2022, {astropy/photutils: 1.5.0}. Zenodo

\bibitem[{{Bravo} {et~al}\mbox{.}(2022){Bravo}, {Piersanti}, {Blondin},
  {Dom{\'\i}nguez}, {Straniero}, \& {Cristallo}}]{bravo22}
{Bravo} E., {Piersanti} L., {Blondin} S., {Dom{\'\i}nguez} I., {Straniero} O.,
  {Cristallo} S., 2022, MNRAS, 517, L31

\bibitem[{{Buccheri} {et~al}\mbox{.}(1983){Buccheri}, {Bennett}, {Bignami},
  {Bloemen}, {Boriakoff}, {Caraveo}, {Hermsen}, {Kanbach}, {Manchester},
  {Masnou}, {Mayer-Hasselwander}, {Ozel}, {Paul}, {Sacco}, {Scarsi}, \&
  {Strong}}]{buccheri83}
{Buccheri} R. {et~al.}, 1983, A\&A, 128, 245

\bibitem[{{Caiazzo} {et~al}\mbox{.}(2021){Caiazzo}, {Burdge}, {Fuller}, {Heyl},
  {Kulkarni}, {Prince}, {Richer}, {Schwab}, {Andreoni}, {Bellm}, {Drake},
  {Duev}, {Graham}, {Helou}, {Mahabal}, {Masci}, {Smith}, \&
  {Soumagnac}}]{caiazzo21}
{Caiazzo} I. {et~al.}, 2021, Nature, 595, 39

\bibitem[{{Coughenour} {et~al}\mbox{.}(2022){Coughenour}, {Tomsick}, {Shaw},
  {Mukai}, {Clavel}, {Hare}, {Krivonos}, \& {Fornasini}}]{coughenour22}
{Coughenour} B.~M., {Tomsick} J.~A., {Shaw} A.~W., {Mukai} K., {Clavel} M.,
  {Hare} J., {Krivonos} R., {Fornasini} F.~M., 2022, MNRAS, 511, 4582

\bibitem[{{Covington} {et~al}\mbox{.}(2022){Covington}, {Shaw}, {Mukai},
  {Littlefield}, {Heinke}, {Plotkin}, {Barrett}, {Boardman}, {Boyd}, {Brincat},
  {Carstens}, {Collins}, {Cook}, {Cooney}, {Fern{\'a}ndez}, {Dufoer}, {Dvorak},
  {Galdies}, {Goff}, {Hambsch}, {Johnston}, {Jones}, {Menzies}, {Monard},
  {Morelle}, {Nelson}, {{\"O}gmen}, {Rock}, {Sabo}, {Seargeant}, {Stone},
  {Ulowetz}, \& {Vanmunster}}]{covington22}
{Covington} A.~E. {et~al.}, 2022, ApJ, 928, 164

\bibitem[{{de Martino} {et~al}\mbox{.}(2020){de Martino}, {Bernardini},
  {Mukai}, {Falanga}, \& {Masetti}}]{demartino20}
{de Martino} D., {Bernardini} F., {Mukai} K., {Falanga} M., {Masetti} N., 2020,
  Advances in Space Research, 66, 1209

\bibitem[{{Di Stefano}(2010{\natexlab{a}})}]{distefano10a}
{Di Stefano} R., 2010{\natexlab{a}}, ApJ, 712, 728

\bibitem[{{Di Stefano}(2010{\natexlab{b}})}]{distefano10b}
{Di Stefano} R., 2010{\natexlab{b}}, ApJ, 719, 474

\bibitem[{{Fitzpatrick}(1999)}]{fitzpatrick99}
{Fitzpatrick} E.~L., 1999, PASP, 111, 63

\bibitem[{{Fornasini} {et~al}\mbox{.}(2017){Fornasini}, {Tomsick}, {Hong},
  {Gotthelf}, {Bauer}, {Rahoui}, {Stern}, {Bodaghee}, {Chiu}, {Clavel},
  {Corral-Santana}, {Hailey}, {Krivonos}, {Mori}, {Alexander}, {Barret},
  {Boggs}, {Christensen}, {Craig}, {Forster}, {Giommi}, {Grefenstette},
  {Harrison}, {Hornstrup}, {Kitaguchi}, {Koglin}, {Madsen}, {Mao}, {Miyasaka},
  {Perri}, {Pivovaroff}, {Puccetti}, {Rana}, {Westergaard}, \&
  {Zhang}}]{fornasini17}
{Fornasini} F.~M. {et~al.}, 2017, ApJS, 229, 33

\bibitem[{{Gaia Collaboration}(2020)}]{gaia_edr3}
{Gaia Collaboration}, 2020, VizieR Online Data Catalog, I/350

\bibitem[{{Gaia Collaboration} {et~al}\mbox{.}(2022){Gaia Collaboration},
  {Vallenari}, {Brown}, {Prusti}, {de Bruijne}, {Arenou}, {Babusiaux},
  {Biermann}, {Creevey}, {Ducourant}, {Evans}, {Eyer}, {Guerra}, {Hutton},
  {Jordi}, {Klioner}, {Lammers}, {Lindegren}, {Luri}, {Mignard}, {Panem},
  {Pourbaix}, {Randich}, {Sartoretti}, {Soubiran}, {Tanga}, {Walton},
  {Bailer-Jones}, {Bastian}, {Drimmel}, {Jansen}, {Katz}, {Lattanzi}, {van
  Leeuwen}, {Bakker}, {Cacciari}, {Casta{\~n}eda}, {De Angeli}, {Fabricius},
  {Fouesneau}, {Fr{\'e}mat}, {Galluccio}, {Guerrier}, {Heiter}, {Masana},
  {Messineo}, {Mowlavi}, {Nicolas}, {Nienartowicz}, {Pailler}, {Panuzzo},
  {Riclet}, {Roux}, {Seabroke}, {Sordo{\o}rcit}, {Th{\'e}venin},
  {Gracia-Abril}, {Portell}, {Teyssier}, {Altmann}, {Andrae}, {Audard},
  {Bellas-Velidis}, {Benson}, {Berthier}, {Blomme}, {Burgess}, {Busonero},
  {Busso}, {C{\'a}novas}, {Carry}, {Cellino}, {Cheek}, {Clementini},
  {Damerdji}, {Davidson}, {de Teodoro}, {Nu{\~n}ez Campos}, {Delchambre},
  {Dell'Oro}, {Esquej}, {Fern{\'a}ndez-Hern{\'a}ndez}, {Fraile}, {Garabato},
  {Garc{\'\i}a-Lario}, {Gosset}, {Haigron}, {Halbwachs}, {Hambly}, {Harrison},
  {Hern{\'a}ndez}, {Hestroffer}, {Hodgkin}, {Holl}, {Jan{\ss}en}, {Jevardat de
  Fombelle}, {Jordan}, {Krone-Martins}, {Lanzafame}, {L{\"o}ffler}, {Marchal},
  {Marrese}, {Moitinho}, {Muinonen}, {Osborne}, {Pancino}, {Pauwels},
  {Recio-Blanco}, {Reyl{\'e}}, {Riello}, {Rimoldini}, {Roegiers}, {Rybizki},
  {Sarro}, {Siopis}, {Smith}, {Sozzetti}, {Utrilla}, {van Leeuwen}, {Abbas},
  {{\'A}brah{\'a}m}, {Abreu Aramburu}, {Aerts}, {Aguado}, {Ajaj},
  {Aldea-Montero}, {Altavilla}, {{\'A}lvarez}, {Alves}, {Anders}, {Anderson},
  {Anglada Varela}, {Antoja}, {Baines}, {Baker}, {Balaguer-N{\'u}{\~n}ez},
  {Balbinot}, {Balog}, {Barache}, {Barbato}, {Barros}, {Barstow},
  {Bartolom{\'e}}, {Bassilana}, {Bauchet}, {Becciani}, {Bellazzini},
  {Berihuete}, {Bernet}, {Bertone}, {Bianchi}, {Binnenfeld}, {Blanco-Cuaresma},
  {Blazere}, {Boch}, {Bombrun}, {Bossini}, {Bouquillon}, {Bragaglia},
  {Bramante}, {Breedt}, {Bressan}, {Brouillet}, {Brugaletta}, {Bucciarelli},
  {Burlacu}, {Butkevich}, {Buzzi}, {Caffau}, {Cancelliere}, {Cantat-Gaudin},
  {Carballo}, {Carlucci}, {Carnerero}, {Carrasco}, {Casamiquela}, {Castellani},
  {Castro-Ginard}, {Chaoul}, {Charlot}, {Chemin}, {Chiaramida}, {Chiavassa},
  {Chornay}, {Comoretto}, {Contursi}, {Cooper}, {Cornez}, {Cowell}, {Crifo},
  {Cropper}, {Crosta}, {Crowley}, {Dafonte}, {Dapergolas}, {David}, {David},
  {de Laverny}, {De Luise}, {De March}, {De Ridder}, {de Souza}, {de Torres},
  {del Peloso}, {del Pozo}, {Delbo}, {Delgado}, {Delisle}, {Demouchy},
  {Dharmawardena}, {Di Matteo}, {Diakite}, {Diener}, {Distefano}, {Dolding},
  {Edvardsson}, {Enke}, {Fabre}, {Fabrizio}, {Faigler}, {Fedorets}, {Fernique},
  {Fienga}, {Figueras}, {Fournier}, {Fouron}, {Fragkoudi}, {Gai},
  {Garcia-Gutierrez}, {Garcia-Reinaldos}, {Garc{\'\i}a-Torres}, {Garofalo},
  {Gavel}, {Gavras}, {Gerlach}, {Geyer}, {Giacobbe}, {Gilmore}, {Girona},
  {Giuffrida}, {Gomel}, {Gomez}, {Gonz{\'a}lez-N{\'u}{\~n}ez},
  {Gonz{\'a}lez-Santamar{\'\i}a}, {Gonz{\'a}lez-Vidal}, {Granvik}, {Guillout},
  {Guiraud}, {Guti{\'e}rrez-S{\'a}nchez}, {Guy}, {Hatzidimitriou}, {Hauser},
  {Haywood}, {Helmer}, {Helmi}, {Sarmiento}, {Hidalgo}, {Hilger},
  {H{\l}adczuk}, {Hobbs}, {Holland}, {Huckle}, {Jardine}, {Jasniewicz},
  {Jean-Antoine Piccolo}, {Jim{\'e}nez-Arranz}, {Jorissen}, {Juaristi
  Campillo}, {Julbe}, {Karbevska}, {Kervella}, {Khanna}, {Kontizas},
  {Kordopatis}, {Korn}, {K{\'o}sp{\'a}l}, {Kostrzewa-Rutkowska},
  {Kruszy{\'n}ska}, {Kun}, {Laizeau}, {Lambert}, {Lanza}, {Lasne}, {Le
  Campion}, {Lebreton}, {Lebzelter}, {Leccia}, {Leclerc}, {Lecoeur-Taibi},
  {Liao}, {Licata}, {Lindstr{\o}m}, {Lister}, {Livanou}, {Lobel}, {Lorca},
  {Loup}, {Madrero Pardo}, {Magdaleno Romeo}, {Managau}, {Mann}, {Manteiga},
  {Marchant}, {Marconi}, {Marcos}, {Marcos Santos}, {Mar{\'\i}n Pina},
  {Marinoni}, {Marocco}, {Marshall}, {Polo}, {Mart{\'\i}n-Fleitas}, {Marton},
  {Mary}, {Masip}, {Massari}, {Mastrobuono-Battisti}, {Mazeh}, {McMillan},
  {Messina}, {Michalik}, {Millar}, {Mints}, {Molina}, {Molinaro}, {Moln{\'a}r},
  {Monari}, {Mongui{\'o}}, {Montegriffo}, {Montero}, {Mor}, {Mora},
  {Morbidelli}, {Morel}, {Morris}, {Muraveva}, {Murphy}, {Musella}, {Nagy},
  {Noval}, {Oca{\~n}a}, {Ogden}, {Ordenovic}, {Osinde}, {Pagani}, {Pagano},
  {Palaversa}, {Palicio}, {Pallas-Quintela}, {Panahi}, {Payne-Wardenaar},
  {Pe{\~n}alosa Esteller}, {Penttil{\"a}}, {Pichon}, {Piersimoni}, {Pineau},
  {Plachy}, {Plum}, {Poggio}, {Pr{\v{s}}a}, {Pulone}, {Racero}, {Ragaini},
  {Rainer}, {Raiteri}, {Rambaux}, {Ramos}, {Ramos-Lerate}, {Re Fiorentin},
  {Regibo}, {Richards}, {Rios Diaz}, {Ripepi}, {Riva}, {Rix}, {Rixon},
  {Robichon}, {Robin}, {Robin}, {Roelens}, {Rogues}, {Rohrbasser},
  {Romero-G{\'o}mez}, {Rowell}, {Royer}, {Ruz Mieres}, {Rybicki}, {Sadowski},
  {S{\'a}ez N{\'u}{\~n}ez}, {Sagrist{\`a} Sell{\'e}s}, {Sahlmann}, {Salguero},
  {Samaras}, {Sanchez Gimenez}, {Sanna}, {Santove{\~n}a}, {Sarasso},
  {Schultheis}, {Sciacca}, {Segol}, {Segovia}, {S{\'e}gransan}, {Semeux},
  {Shahaf}, {Siddiqui}, {Siebert}, {Siltala}, {Silvelo}, {Slezak}, {Slezak},
  {Smart}, {Snaith}, {Solano}, {Solitro}, {Souami}, {Souchay}, {Spagna},
  {Spina}, {Spoto}, {Steele}, {Steidelm{\"u}ller}, {Stephenson}, {S{\"u}veges},
  {Surdej}, {Szabados}, {Szegedi-Elek}, {Taris}, {Taylo}, {Teixeira},
  {Tolomei}, {Tonello}, {Torra}, {Torra}, {Torralba Elipe}, {Trabucchi},
  {Tsounis}, {Turon}, {Ulla}, {Unger}, {Vaillant}, {van Dillen}, {van Reeven},
  {Vanel}, {Vecchiato}, {Viala}, {Vicente}, {Voutsinas}, {Weiler}, {Wevers},
  {Wyrzykowski}, {Yoldas}, {Yvard}, {Zhao}, {Zorec}, {Zucker}, \&
  {Zwitter}}]{gaia_dr3}
{Gaia Collaboration} {et~al.}, 2022, arXiv e-prints, arXiv:2208.00211

\bibitem[{{Gilfanov} \& {Bogd{\'a}n}(2010)}]{gb10}
{Gilfanov} M., {Bogd{\'a}n} {\'A}., 2010, Nature, 463, 924

\bibitem[{{Graur} {et~al}\mbox{.}(2014){Graur}, {Maoz}, \& {Shara}}]{graur14}
{Graur} O., {Maoz} D., {Shara} M.~M., 2014, MNRAS, 442, L28

\bibitem[{{G{\"u}ver} \& {{\"O}zel}(2009)}]{go09}
{G{\"u}ver} T., {{\"O}zel} F., 2009, MNRAS, 400, 2050

\bibitem[{{Hailey} {et~al}\mbox{.}(2016){Hailey}, {Mori}, {Perez}, {Canipe},
  {Hong}, {Tomsick}, {Boggs}, {Christensen}, {Craig}, {Fornasini}, {Grindlay},
  {Harrison}, {Nynka}, {Rahoui}, {Stern}, {Zhang}, \& {Zhang}}]{hailey16}
{Hailey} C.~J. {et~al.}, 2016, ApJ, 826, 160

\bibitem[{{Harrison} {et~al}\mbox{.}(2013){Harrison}, {Craig}, {Christensen},
  {Hailey}, {Zhang}, {Boggs}, {Stern}, {Cook}, {Forster}, {Giommi},
  {Grefenstette}, {Kim}, {Kitaguchi}, {Koglin}, {Madsen}, {Mao}, {Miyasaka},
  {Mori}, {Perri}, {Pivovaroff}, {Puccetti}, {Rana}, {Westergaard}, {Willis},
  {Zoglauer}, {An}, {Bachetti}, {Barri{\`e}re}, {Bellm}, {Bhalerao},
  {Brejnholt}, {Fuerst}, {Liebe}, {Markwardt}, {Nynka}, {Vogel}, {Walton},
  {Wik}, {Alexander}, {Cominsky}, {Hornschemeier}, {Hornstrup}, {Kaspi},
  {Madejski}, {Matt}, {Molendi}, {Smith}, {Tomsick}, {Ajello}, {Ballantyne},
  {Balokovi{\'c}}, {Barret}, {Bauer}, {Blandford}, {Brandt}, {Brenneman},
  {Chiang}, {Chakrabarty}, {Chenevez}, {Comastri}, {Dufour}, {Elvis}, {Fabian},
  {Farrah}, {Fryer}, {Gotthelf}, {Grindlay}, {Helfand}, {Krivonos}, {Meier},
  {Miller}, {Natalucci}, {Ogle}, {Ofek}, {Ptak}, {Reynolds}, {Rigby},
  {Tagliaferri}, {Thorsett}, {Treister}, \& {Urry}}]{harrison13}
{Harrison} F.~A. {et~al.}, 2013, ApJ, 770, 103

\bibitem[{{Hayashi} {et~al}\mbox{.}(2011){Hayashi}, {Ishida}, {Terada},
  {Bamba}, \& {Shionome}}]{hayashi11}
{Hayashi} T., {Ishida} M., {Terada} Y., {Bamba} A., {Shionome} T., 2011, PASJ,
  63, S739

\bibitem[{{Hellier} \& {Mukai}(2004)}]{hm04}
{Hellier} C., {Mukai} K., 2004, MNRAS, 352, 1037

\bibitem[{{Henden} {et~al}\mbox{.}(2016){Henden}, {Templeton}, {Terrell},
  {Smith}, {Levine}, \& {Welch}}]{henden16}
{Henden} A.~A., {Templeton} M., {Terrell} D., {Smith} T.~C., {Levine} S.,
  {Welch} D., 2016, VizieR Online Data Catalog, II/336

\bibitem[{{Hong} {et~al}\mbox{.}(2016){Hong}, {Mori}, {Hailey}, {Nynka},
  {Zhang}, {Gotthelf}, {Fornasini}, {Krivonos}, {Bauer}, {Perez}, {Tomsick},
  {Bodaghee}, {Chiu}, {Clavel}, {Stern}, {Grindlay}, {Alexander}, {Aramaki},
  {Baganoff}, {Barret}, {Barri{\`e}re}, {Boggs}, {Canipe}, {Christensen},
  {Craig}, {Desai}, {Forster}, {Giommi}, {Grefenstette}, {Harrison}, {Hong},
  {Hornstrup}, {Kitaguchi}, {Koglin}, {Madsen}, {Mao}, {Miyasaka}, {Perri},
  {Pivovaroff}, {Puccetti}, {Rana}, {Westergaard}, {Zhang}, \&
  {Zoglauer}}]{hong16}
{Hong} J. {et~al.}, 2016, ApJ, 825, 132

\bibitem[{{Kalberla} {et~al}\mbox{.}(2005){Kalberla}, {Burton}, {Hartmann},
  {Arnal}, {Bajaja}, {Morras}, \& {P{\"o}ppel}}]{kalberla05}
{Kalberla} P.~M.~W., {Burton} W.~B., {Hartmann} D., {Arnal} E.~M., {Bajaja} E.,
  {Morras} R., {P{\"o}ppel} W.~G.~L., 2005, A\&A, 440, 775

\bibitem[{{Kilic} {et~al}\mbox{.}(2021){Kilic}, {Kosakowski}, {Moss},
  {Bergeron}, \& {Conly}}]{kilic21}
{Kilic} M., {Kosakowski} A., {Moss} A.~G., {Bergeron} P., {Conly} A.~A., 2021,
  ApJ, 923, L6

\bibitem[{{Krivonos} {et~al}\mbox{.}(2007){Krivonos}, {Revnivtsev}, {Churazov},
  {Sazonov}, {Grebenev}, \& {Sunyaev}}]{krivonos07_cv}
{Krivonos} R., {Revnivtsev} M., {Churazov} E., {Sazonov} S., {Grebenev} S.,
  {Sunyaev} R., 2007, A\&A, 463, 957

\bibitem[{{Krivonos} {et~al}\mbox{.}(2021){Krivonos}, {Bird}, {Churazov},
  {Tomsick}, {Bazzano}, {Beckmann}, {B{\'e}langer}, {Bodaghee}, {Chaty},
  {Kuulkers}, {Lutovinov}, {Malizia}, {Masetti}, {Mereminskiy}, {Sunyaev},
  {Tsygankov}, {Ubertini}, \& {Winkler}}]{krivonos21}
{Krivonos} R.~A. {et~al.}, 2021, New Astronomy Reviews, 92, 101612

\bibitem[{{Krivonos} {et~al}\mbox{.}(2022){Krivonos}, {Sazonov}, {Kuznetsova},
  {Lutovinov}, {Mereminskiy}, \& {Tsygankov}}]{krivonos22}
{Krivonos} R.~A., {Sazonov} S.~Y., {Kuznetsova} E.~A., {Lutovinov} A.~A.,
  {Mereminskiy} I.~A., {Tsygankov} S.~S., 2022, MNRAS, 510, 4796

\bibitem[{{Krivonos} {et~al}\mbox{.}(2017){Krivonos}, {Tsygankov},
  {Mereminskiy}, {Lutovinov}, {Sazonov}, \& {Sunyaev}}]{krivonos17}
{Krivonos} R.~A., {Tsygankov} S.~S., {Mereminskiy} I.~A., {Lutovinov} A.~A.,
  {Sazonov} S.~Y., {Sunyaev} R.~A., 2017, MNRAS, 470, 512

\bibitem[{{Landi} {et~al}\mbox{.}(2017){Landi}, {Bassani}, {Bazzano}, {Bird},
  {Fiocchi}, {Malizia}, {Panessa}, {Sguera}, \& {Ubertini}}]{landi17}
{Landi} R. {et~al.}, 2017, MNRAS, 470, 1107

\bibitem[{{Lang} {et~al}\mbox{.}(2010){Lang}, {Hogg}, {Mierle}, {Blanton}, \&
  {Roweis}}]{lang10}
{Lang} D., {Hogg} D.~W., {Mierle} K., {Blanton} M., {Roweis} S., 2010, AJ, 139,
  1782

\bibitem[{{Leising}(2022)}]{leising22}
{Leising} M.~D., 2022, ApJ, 932, 63

\bibitem[{{Liu} {et~al}\mbox{.}(2012){Liu}, {Di Stefano}, {Wang}, \&
  {Moe}}]{liu12}
{Liu} J., {Di Stefano} R., {Wang} T., {Moe} M., 2012, ApJ, 749, 141

\bibitem[{{Lomb}(1976)}]{lomb76}
{Lomb} N.~R., 1976, Ap\&SS, 39, 447

\bibitem[{{Lutovinov} {et~al}\mbox{.}(2020){Lutovinov}, {Suleimanov}, {Manuel
  Luna}, {Sazonov}, {de Martino}, {Ducci}, {Doroshenko}, \&
  {Falanga}}]{lutovinov20}
{Lutovinov} A., {Suleimanov} V., {Manuel Luna} G.~J., {Sazonov} S., {de
  Martino} D., {Ducci} L., {Doroshenko} V., {Falanga} M., 2020, New Astronomy
  Reviews, 91, 101547

\bibitem[{{Magdziarz} \& {Zdziarski}(1995)}]{mz95}
{Magdziarz} P., {Zdziarski} A.~A., 1995, MNRAS, 273, 837

\bibitem[{{McCully} {et~al}\mbox{.}(2018){McCully}, {Volgenau}, {Harbeck},
  {Lister}, {Saunders}, {Turner}, {Siiverd}, \& {Bowman}}]{mccully18}
{McCully} C., {Volgenau} N.~H., {Harbeck} D.-R., {Lister} T.~A., {Saunders}
  E.~S., {Turner} M.~L., {Siiverd} R.~J., {Bowman} M., 2018, in Society of
  Photo-Optical Instrumentation Engineers (SPIE) Conference Series, Vol. 10707,
  Software and Cyberinfrastructure for Astronomy V, {Guzman} J.~C., {Ibsen} J.,
  eds., p. 107070K

\bibitem[{{Miller} {et~al}\mbox{.}(2022){Miller}, {Caiazzo}, {Heyl}, {Richer},
  \& {Tremblay}}]{miller22}
{Miller} D.~R., {Caiazzo} I., {Heyl} J., {Richer} H.~B., {Tremblay} P.-E.,
  2022, ApJ, 926, L24

\bibitem[{{Mukai} {et~al}\mbox{.}(2015){Mukai}, {Rana}, {Bernardini}, \& {de
  Martino}}]{mukai15}
{Mukai} K., {Rana} V., {Bernardini} F., {de Martino} D., 2015, ApJ, 807, L30

\bibitem[{{Neopane} {et~al}\mbox{.}(2022){Neopane}, {Bhargava}, {Fisher},
  {Ferrari}, {Yoshida}, {Toonen}, \& {Bravo}}]{neopane22}
{Neopane} S., {Bhargava} K., {Fisher} R., {Ferrari} M., {Yoshida} S., {Toonen}
  S., {Bravo} E., 2022, ApJ, 925, 92

\bibitem[{{Nielsen} {et~al}\mbox{.}(2013{\natexlab{a}}){Nielsen}, {Dominik},
  {Nelemans}, \& {Voss}}]{nielsen13a}
{Nielsen} M.~T.~B., {Dominik} C., {Nelemans} G., {Voss} R., 2013{\natexlab{a}},
  A\&A, 549, A32

\bibitem[{{Nielsen} {et~al}\mbox{.}(2014){Nielsen}, {Gilfanov}, {Bogd{\'a}n},
  {Woods}, \& {Nelemans}}]{nielsen14}
{Nielsen} M.~T.~B., {Gilfanov} M., {Bogd{\'a}n} {\'A}., {Woods} T.~E.,
  {Nelemans} G., 2014, MNRAS, 442, 3400

\bibitem[{{Nielsen} {et~al}\mbox{.}(2012){Nielsen}, {Voss}, \&
  {Nelemans}}]{nielsen12}
{Nielsen} M.~T.~B., {Voss} R., {Nelemans} G., 2012, MNRAS, 426, 2668

\bibitem[{{Nielsen} {et~al}\mbox{.}(2013{\natexlab{b}}){Nielsen}, {Voss}, \&
  {Nelemans}}]{nielsen13b}
{Nielsen} M.~T.~B., {Voss} R., {Nelemans} G., 2013{\natexlab{b}}, MNRAS, 435,
  187

\bibitem[{{Pecaut} \& {Mamajek}(2013)}]{pm13}
{Pecaut} M.~J., {Mamajek} E.~E., 2013, ApJS, 208, 9

\bibitem[{{Perez} {et~al}\mbox{.}(2015){Perez}, {Hailey}, {Bauer}, {Krivonos},
  {Mori}, {Baganoff}, {Barri{\`e}re}, {Boggs}, {Christensen}, {Craig},
  {Grefenstette}, {Grindlay}, {Harrison}, {Hong}, {Madsen}, {Nynka}, {Stern},
  {Tomsick}, {Wik}, {Zhang}, {Zhang}, \& {Zoglauer}}]{perez15}
{Perez} K. {et~al.}, 2015, Nature, 520, 646

\bibitem[{{Ritter} \& {Kolb}(2011)}]{rk11}
{Ritter} H., {Kolb} U., 2011, VizieR Online Data Catalog, B/cb

\bibitem[{{Scargle}(1982)}]{scargle82}
{Scargle} J.~D., 1982, ApJ, 263, 835

\bibitem[{{Shaw} {et~al}\mbox{.}(2018){Shaw}, {Heinke}, {Mukai}, {Sivakoff},
  {Tomsick}, \& {Rana}}]{shaw18}
{Shaw} A.~W., {Heinke} C.~O., {Mukai} K., {Sivakoff} G.~R., {Tomsick} J.~A.,
  {Rana} V., 2018, MNRAS, 476, 554

\bibitem[{{Shaw} {et~al}\mbox{.}(2020){Shaw}, {Heinke}, {Mukai}, {Tomsick},
  {Doroshenko}, {Suleimanov}, {Buisson}, {Gandhi}, {Grefenstette}, {Hare},
  {Jiang}, {Ludlam}, {Rana}, \& {Sivakoff}}]{shaw20}
{Shaw} A.~W. {et~al.}, 2020, MNRAS, 498, 3457

\bibitem[{{Starrfield} {et~al}\mbox{.}(2020){Starrfield}, {Bose}, {Iliadis},
  {Hix}, {Woodward}, \& {Wagner}}]{starrfield20}
{Starrfield} S., {Bose} M., {Iliadis} C., {Hix} W.~R., {Woodward} C.~E.,
  {Wagner} R.~M., 2020, ApJ, 895, 70

\bibitem[{{Str{\"u}der} {et~al}\mbox{.}(2001){Str{\"u}der}, {Briel}, {Dennerl},
  {Hartmann}, {Kendziorra}, {Meidinger}, {Pfeffermann}, {Reppin}, {Aschenbach},
  {Bornemann}, {Br{\"a}uninger}, {Burkert}, {Elender}, {Freyberg}, {Haberl},
  {Hartner}, {Heuschmann}, {Hippmann}, {Kastelic}, {Kemmer}, {Kettenring},
  {Kink}, {Krause}, {M{\"u}ller}, {Oppitz}, {Pietsch}, {Popp}, {Predehl},
  {Read}, {Stephan}, {St{\"o}tter}, {Tr{\"u}mper}, {Holl}, {Kemmer}, {Soltau},
  {St{\"o}tter}, {Weber}, {Weichert}, {von Zanthier}, {Carathanassis}, {Lutz},
  {Richter}, {Solc}, {B{\"o}ttcher}, {Kuster}, {Staubert}, {Abbey}, {Holland},
  {Turner}, {Balasini}, {Bignami}, {La Palombara}, {Villa}, {Buttler},
  {Gianini}, {Lain{\'e}}, {Lumb}, \& {Dhez}}]{struder01}
{Str{\"u}der} L. {et~al.}, 2001, A\&A, 365, L18

\bibitem[{{Suleimanov} {et~al}\mbox{.}(2016){Suleimanov}, {Doroshenko},
  {Ducci}, {Zhukov}, \& {Werner}}]{suleimanov16}
{Suleimanov} V., {Doroshenko} V., {Ducci} L., {Zhukov} G.~V., {Werner} K.,
  2016, A\&A, 591, A35

\bibitem[{{Suleimanov} {et~al}\mbox{.}(2005){Suleimanov}, {Revnivtsev}, \&
  {Ritter}}]{srr05}
{Suleimanov} V., {Revnivtsev} M., {Ritter} H., 2005, A\&A, 435, 191

\bibitem[{{Suleimanov} {et~al}\mbox{.}(2019){Suleimanov}, {Doroshenko}, \&
  {Werner}}]{suleimanov19}
{Suleimanov} V.~F., {Doroshenko} V., {Werner} K., 2019, MNRAS, 482, 3622

\bibitem[{{Suleimanov} {et~al}\mbox{.}(2022){Suleimanov}, {Doroshenko}, \&
  {Werner}}]{suleimanov22}
{Suleimanov} V.~F., {Doroshenko} V., {Werner} K., 2022, \mnras, 511, 4937

\bibitem[{{Taylor} {et~al}\mbox{.}(1997){Taylor}, {Beardmore}, {Norton},
  {Osborne}, \& {Watson}}]{taylor97}
{Taylor} P., {Beardmore} A.~P., {Norton} A.~J., {Osborne} J.~P., {Watson}
  M.~G., 1997, MNRAS, 289, 349

\bibitem[{{Tomsick} {et~al}\mbox{.}(2020){Tomsick}, {Bodaghee}, {Chaty},
  {Clavel}, {Fornasini}, {Hare}, {Krivonos}, {Rahoui}, \&
  {Rodriguez}}]{tomsick20}
{Tomsick} J.~A. {et~al.}, 2020, ApJ, 889, 53

\bibitem[{{Tomsick} {et~al}\mbox{.}(2016){Tomsick}, {Rahoui}, {Krivonos},
  {Clavel}, {Strader}, \& {Chomiuk}}]{tomsick16b}
{Tomsick} J.~A., {Rahoui} F., {Krivonos} R., {Clavel} M., {Strader} J.,
  {Chomiuk} L., 2016, MNRAS, 460, 513

\bibitem[{{Turner} {et~al}\mbox{.}(2001){Turner}, {Abbey}, {Arnaud},
  {Balasini}, {Barbera}, {Belsole}, {Bennie}, {Bernard}, {Bignami}, {Boer},
  {Briel}, {Butler}, {Cara}, {Chabaud}, {Cole}, {Collura}, {Conte}, {Cros},
  {Denby}, {Dhez}, {Di Coco}, {Dowson}, {Ferrando}, {Ghizzardi}, {Gianotti},
  {Goodall}, {Gretton}, {Griffiths}, {Hainaut}, {Hochedez}, {Holland},
  {Jourdain}, {Kendziorra}, {Lagostina}, {Laine}, {La Palombara}, {Lortholary},
  {Lumb}, {Marty}, {Molendi}, {Pigot}, {Poindron}, {Pounds}, {Reeves},
  {Reppin}, {Rothenflug}, {Salvetat}, {Sauvageot}, {Schmitt}, {Sembay},
  {Short}, {Spragg}, {Stephen}, {Str{\"u}der}, {Tiengo}, {Trifoglio},
  {Tr{\"u}mper}, {Vercellone}, {Vigroux}, {Villa}, {Ward}, {Whitehead}, \&
  {Zonca}}]{turner01}
{Turner} M.~J.~L. {et~al.}, 2001, A\&A, 365, L27

\bibitem[{{Verner} {et~al}\mbox{.}(1996){Verner}, {Ferland}, {Korista}, \&
  {Yakovlev}}]{vern96}
{Verner} D.~A., {Ferland} G.~J., {Korista} K.~T., {Yakovlev} D.~G., 1996, ApJ,
  465, 487

\bibitem[{{Wilms} {et~al}\mbox{.}(2000){Wilms}, {Allen}, \& {McCray}}]{wam00}
{Wilms} J., {Allen} A., {McCray} R., 2000, ApJ, 542, 914

\bibitem[{{Xu} {et~al}\mbox{.}(2019){Xu}, {Yu}, \& {Li}}]{xu19}
{Xu} X.-j., {Yu} Z.-l., {Li} X.-d., 2019, ApJ, 878, 53

\bibitem[{{Yu} {et~al}\mbox{.}(2022){Yu}, {Xu}, \& {Li}}]{yu22}
{Yu} Z.-L., {Xu} X.-J., {Li} X.-D., 2022, Research in Astronomy and
  Astrophysics, 22, 045003

\bibitem[{{Yuasa} {et~al}\mbox{.}(2012){Yuasa}, {Makishima}, \&
  {Nakazawa}}]{yuasa12}
{Yuasa} T., {Makishima} K., {Nakazawa} K., 2012, ApJ, 753, 129

\bibitem[{{Zorotovic} {et~al}\mbox{.}(2011){Zorotovic}, {Schreiber}, \&
  {G{\"a}nsicke}}]{zsg11}
{Zorotovic} M., {Schreiber} M.~R., {G{\"a}nsicke} B.~T., 2011, A\&A, 536, A42

\end{thebibliography}


\section*{Appendix A: XMMU~J150407.8--602227}

\begin{figure*}
\begin{center}
    \includegraphics[width=15cm]{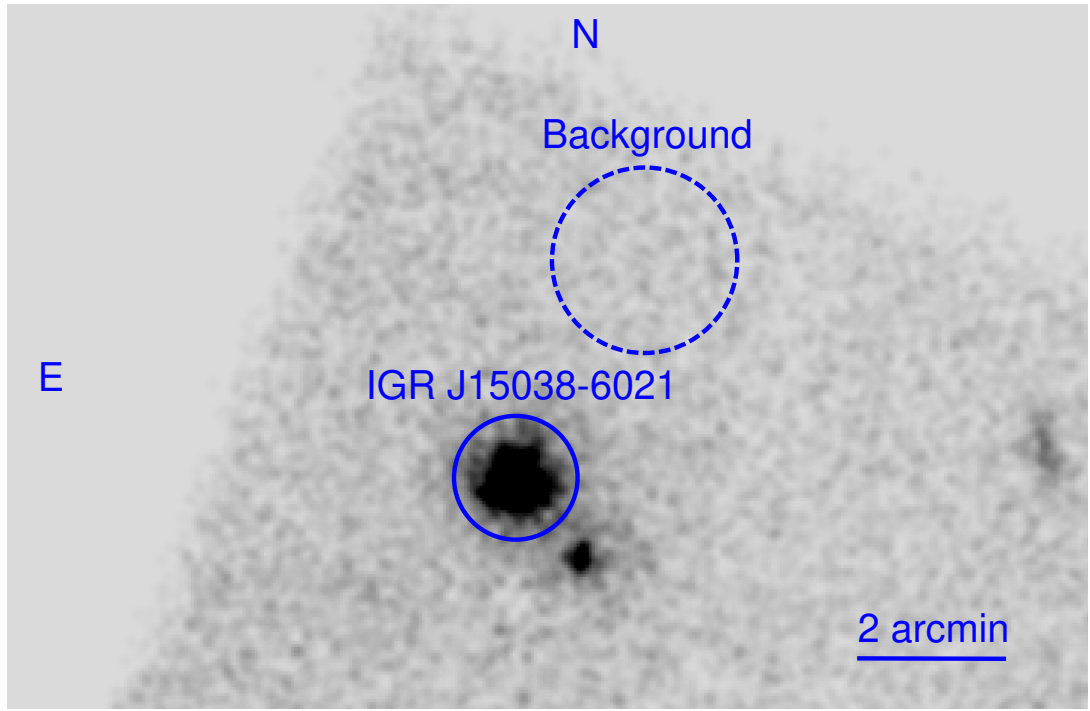}
    \caption{The 3-79\,keV {\em NuSTAR} FPMA image.  The J15038 source region is shown as a solid circle with a radius of $50''$.  The background region used is shown as a dashed circle with a radius of $75''$.  J15038 is clearly separated from the nearby source to the Southwest (XMMU~J150407.8--602227).  There is another serendipitous source detected at West edge of the image.
    \label{fig:image}}
\end{center}
\end{figure*}

Although considerably fainter than J15038, we performed some additional analysis of the nearby source XMMU~J150407.8--602227, which was detected by {\em XMM} and {\em NuSTAR} (see Figure~\ref{fig:image}).  The {\em XMM}/pn+{\em NuSTAR} spectrum can be reasonably well-fit ($\chi^{2}/$dof = 53/44 with an absorbed thermal bremsstrahlung model with $N_{\rm H} = (3.8\pm 0.7)\times 10^{22}$\,cm$^{-2}$ and $kT = 25^{+11}_{-7}$\,keV.  The source has a 0.3-79\,keV absorbed flux of $1.2\times 10^{-12}$\,erg\,cm$^{-2}$\,s$^{-1}$.  There are significant positive residuals in the 6-7\,keV iron line region, which may indicate the presence of an iron line.  

The J15038 field was previously observed with {\em Chandra} on 2017 April 26 with a 5\,ks exposure \citep{tomsick20}, and XMMU~J150407.8--602227 is not detected.  If the source was at the flux seen by {\em XMM}+{\em NuSTAR}, approximately 96 counts would have been detected in {\em Chandra}/ACIS, indicating that the source flux changed between 2017 April and 2020 July by a factor of 20 or more.  Such a high level of variability rules out an AGN and disfavors an IP classification.  However, an IP cannot be completely ruled out since low flux states are sometimes seen for IPs \citep[e.g.][]{covington22}.

The best {\em XMM} position estimate is $1.2''$ away from the VISTA source VVV J150407.69--602226.88 with $Z=19.51\pm 0.13$, $Y=18.81\pm 0.12$, $J=18.05\pm 0.09$, $H=17.40\pm 0.09$, and $K_{s}=17.14\pm 0.13$. There is no optical source present in the LCO $r'$-band data. The positions are close enough for the VISTA source to be the counterpart, and it may be, but if the column density of $3.8\times 10^{22}$\,cm$^{-2}$ is interstellar, then one would expect the VISTA colors to show much higher extinction.  Although there is not enough information for a definite classification, the absorption may be local to the source as seen, for example, for obscured HMXBs.

\label{lastpage}
\bsp


\begin{figure*}
\begin{center}
\includegraphics[width=15cm,angle=0]{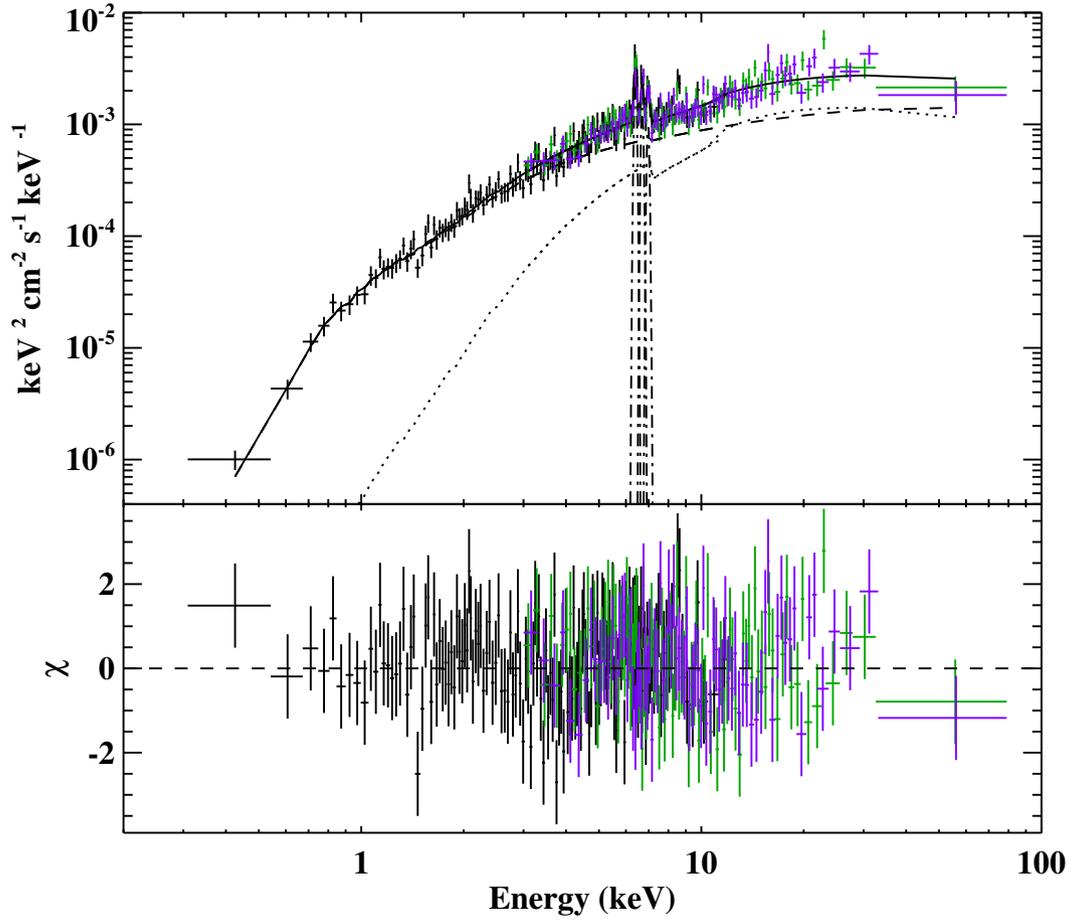}
\caption{Unfolded {\em XMM} and {\em NuSTAR} energy spectrum fitted with the PSR model fit shown in Table~\ref{tab:spec2}.  The components are three Gaussians {\em (dash-dotted lines)}, reflection {\em (dotted line)}, and the PSR model {\em (dashed)}.  We fitted the data with all three {\em XMM} instruments (pn, MOS1, and MOS2), but, for clarity, we only show the pn data.\label{fig:spectrum}}
\end{center}
\end{figure*}

\clearpage


\begin{table*}
\caption{Observations of IGR~J15038--6021\label{tab:obs}}
\begin{minipage}{\linewidth}
\begin{tabular}{cccccccc} \hline \hline
Observatory  & ObsID         & Instrument & Start Time (UT) & End Time (UT) & Exposure Time (ks)\\ \hline\hline
{\em XMM}    & 0870790101    & pn         & 2020 July 30, 11.63 h & 2020 July 30, 18.51 h & 20.3\\
''           & ''            & MOS1       & 2020 July 30, 11.21 h & 2020 July 30, 18.60 h & 25.0\\
''           & ''            & MOS2       & ''                    & ''                    & 25.0\\
{\em NuSTAR} & 30601016002   & FPMA       & 2020 July 30, 10.44 h & 2020 July 31, 10.35 h & 43.2\\
''           & ''            & FPMB       & ''                    & ''                    & 42.9\\
 \hline
\end{tabular}
\end{minipage}
\end{table*}

\begin{table*}
\caption{Quality of fits to {\em XMM}+{\em NuSTAR} energy spectra\label{tab:quality}}
\begin{minipage}{\linewidth}
\begin{tabular}{lc} \hline \hline
Model\footnote{in XSPEC notation.} & ${\chi}^2$/dof\\ \hline
\ttfamily{constant*tbabs*pegpwrlw} & 1333/582\\ \hline
\ttfamily{constant*tbabs(gaussian+pegpwrlw)} & 833/579\\ \hline
\ttfamily{constant*tbabs(gaussian+bremss)} & 1020/579\\ \hline
\ttfamily{constant*tbabs*pcfabs(gaussian+bremss)} & 568/577\\ \hline
\ttfamily{constant*tbabs*pcfabs(gaussian+reflect*bremss)} & 541/575\\ \hline
\ttfamily{constant*tbabs*pcfabs(gaussian+gaussian+gaussian+reflect*bremss)} & 542/575\\ \hline
\ttfamily{constant*tbabs*pcfabs(gaussian+gaussian+gaussian+reflect*atable\{ipolar.fits\})} & 543/575\\ \hline
\end{tabular}
\end{minipage}
\end{table*}

\begin{table*}
\caption{Spectral Results for Bremsstrahlung Fits\label{tab:spec1}}
\begin{minipage}{\linewidth}
\begin{tabular}{cccc} \hline \hline
Parameter\footnote{The errors on the parameters are 90\% confidence.} & Units & 1 Gaussian\footnote{The full model in XSPEC is {\ttfamily constant*tbabs*pcfabs*(gaussian+reflect*bremss)}.} & 3 Gaussians\footnote{This is the same model as the first column except for two additional Gaussians.  Ditto marks ('') indicate parameters that are consistent with the 1 Gaussian column.}\\ \hline
$N_{\rm H}$\footnote{The column density is calculated assuming \cite{wam00} abundances and \cite{vern96} cross sections.  Along this line of sight, the Galactic value is $N_{\rm H} = 1.8\times 10^{22}$\,cm$^{-2}$ \citep{kalberla05}.} & $10^{22}$\,cm$^{-2}$ & $0.43^{+0.07}_{-0.06}$ & ''\\
$N_{\rm H,pc}$ & $10^{22}$\,cm$^{-2}$ & $5.9^{+1.8}_{-1.6}$ & ''\\
pc fraction & --- & $0.68^{+0.05}_{-0.10}$ & ''\\ \hline
$kT$ & keV & $57^{+39}_{-18}$ & ''\\
$N_{\rm bremss}$\footnote{The normalization for the {\ttfamily bremss} model is equal to $\frac{3.02\times 10^{-15}}{4\pi d^{2}}\int{n_{e} n_{i} dV}$, where $d$ is the distance to the source in units of cm, $n_{e}$ and $n_{i}$ are the electron and ion densities in the plasma, and $V$ is the volume of the region containing the plasma.} & --- & $(4.6^{+0.7}_{-1.0})\times 10^{-4}$ & ''\\ \hline
$\Omega/2\pi$ & --- & 1.0\footnote{Fixed.} & ''\\
$A$\footnote{The abundance of elements heavier than helium relative to solar.} & --- & $0.28^{+0.58}_{-0.24}$ & ''\\
$A_{\rm Fe}$\footnote{The abundance of iron relative to solar.} & --- & 0.28\footnote{Tied to $A$.} & ''\\
$\cos{i}$ & --- & $0.57^{+0.38}_{-0.25}$ & ''\\ \hline
$E_{\rm line1}$ & keV & $6.52\pm 0.05$ & 6.4$^{f}$\\
$\sigma_{\rm line1}$ & keV & $0.32^{+0.06}_{-0.05}$ & 0.05$^{f}$\\
$N_{\rm line1}$ & ph\,cm$^{-2}$\,s$^{-1}$ & $(1.9\pm 0.3)\times 10^{-5}$ & $(8.7\pm 1.4)\times 10^{-6}$\\
$EW_{\rm line1}$ & eV & $720\pm 102$ & $273\pm 44$\\ \hline
$E_{\rm line2}$ & keV & --- & 6.7$^{f}$\\
$\sigma_{\rm line2}$ & keV & --- & 0.05$^{f}$\\
$N_{\rm line2}$ & ph\,cm$^{-2}$\,s$^{-1}$ & --- & $(4.4\pm 1.4)\times 10^{-6}$\\
$EW_{\rm line2}$ & eV & --- &  $109\pm 35$\\ \hline
$E_{\rm line3}$ & keV & --- & 6.97$^{f}$\\
$\sigma_{\rm line3}$ & keV & --- & 0.05$^{f}$\\
$N_{\rm line3}$ & ph\,cm$^{-2}$\,s$^{-1}$ & --- & $(3.2\pm 1.3)\times 10^{-6}$\\
$EW_{\rm line3}$ & eV & --- & $104\pm 43$\\ \hline
$C_{\rm pn}$ & --- & 1.0$^{f}$ & ''\\
$C_{\rm MOS1}$ & --- & $1.01\pm 0.05$ & ''\\
$C_{\rm MOS2}$ & --- & $1.00\pm 0.05$ & ''\\
$C_{\rm FPMA}$ & --- & $1.21\pm 0.06$ & ''\\
$C_{\rm FPMB}$ & --- & $1.20\pm 0.06$ & ''\\ \hline
$\chi^{2}$/dof & --- & 541/575 & 542/575\\ \hline
\end{tabular}
\end{minipage}
\end{table*}

\begin{table*}
\caption{Spectral Results for PSR Model Fits\label{tab:spec2}}
\begin{minipage}{\linewidth}
\begin{tabular}{ccc} \hline \hline
Parameter\footnote{The errors on the parameters are 90\% confidence.} & Units & PSR\footnote{This is the model developed by \cite{suleimanov16} for IPs.  The full model is {\ttfamily constant*tbabs*pcfabs*(gaussian+gaussian+gaussian + reflect*atable\{ipolar.fits\})}.  PSR refers to the XSPEC table model called {\ttfamily ipolar.fits}.} model parameters\\ \hline
$N_{\rm H}$\footnote{The column density is calculated assuming \cite{wam00} abundances and \cite{vern96} cross sections.  Along this line of sight, the Galactic value is $N_{\rm H} = 1.8\times 10^{22}$\,cm$^{-2}$ \citep{kalberla05}.} & $10^{22}$\,cm$^{-2}$ & $0.45^{+0.07}_{-0.06}$\\
$N_{\rm H,pc}$ & $10^{22}$\,cm$^{-2}$ & $6.1^{+1.8}_{-1.3}$\\
pc fraction & --- & $0.68^{+0.05}_{-0.08}$\\ \hline
$M_{\rm WD}$ & \Msun & $1.36^{+0.04}_{-0.11}$\\
$R_{\rm m}$\footnote{The radius where the accretion disk is magnetically disrupted and free-fall begins.  The calculated value uses the WD spin rate of 1678\,s that we obtain in Section 3.1 of this work.} & $R_{\rm WD}$ & 107\\
$N_{\rm PSR}$ & --- & $(4.8^{+13}_{-3.3})\times 10^{-30}$\\ \hline
$\Omega/2\pi$ & --- & 1.0\footnote{Fixed.}\\
$A$\footnote{The abundance of elements heavier than helium relative to solar.} & --- & $0.15^{+0.29}_{-0.09}$\\
$A_{\rm Fe}$\footnote{The abundance of iron relative to solar.} & --- & 0.15\footnote{Tied to $A$.}\\
$\cos{i}$ & --- & $0.80^{+0.15}_{-0.31}$\\ \hline
$E_{\rm line1}$ & keV & 6.4$^{e}$\\
$\sigma_{\rm line1}$ & keV & 0.05$^{e}$\\
$N_{\rm line1}$ & ph\,cm$^{-2}$\,s$^{-1}$ & $(8.7\pm 1.4)\times 10^{-6}$\\ \hline
$E_{\rm line2}$ & keV & 6.7$^{e}$\\
$\sigma_{\rm line2}$ & keV & 0.05$^{e}$\\
$N_{\rm line2}$ & ph\,cm$^{-2}$\,s$^{-1}$ & $(4.4\pm 1.4)\times 10^{-6}$\\ \hline
$E_{\rm line3}$ & keV & 6.97$^{e}$\\
$\sigma_{\rm line3}$ & keV & $0.05^{e}$\\
$N_{\rm line3}$ & ph\,cm$^{-2}$\,s$^{-1}$ & $(3.2\pm 1.3)\times 10^{-6}$\\ \hline
$C_{\rm pn}$ & --- & 1.0$^{e}$\\
$C_{\rm MOS1}$ & --- & $1.01\pm 0.05$\\
$C_{\rm MOS2}$ & --- & $1.00\pm 0.05$\\
$C_{\rm FPMA}$ & --- & $1.20\pm 0.06$\\
$C_{\rm FPMB}$ & --- & $1.19\pm 0.06$\\ \hline
$\chi^{2}$/dof & --- & 543/575\\ \hline
\end{tabular}
\end{minipage}
\end{table*}

\end{document}